\documentclass[10pt,aps,pra,twocolumn,reprint,noeprint]{revtex4-1}
\usepackage{amsmath,amsfonts,amssymb,bbm}
\usepackage{graphicx}
\usepackage{xcolor}
\usepackage{soul}

\usepackage{tikz}

\newcommand{\vstrut}{\rule[-2mm]{0mm}{6.5mm}}
\newcommand{\yes}{$\bullet$}
\newcommand{\no}{--}

\begin{document}

\title{Interacting Majorana modes at surfaces of noncentrosymmetric superconductors}

\author{Janna E. R\"uckert}
\affiliation{Institute of Theoretical Physics, Technische
  Universit\"at Dresden, 01062 Dresden, Germany}
\author{Gerg\H{o} Ro\'osz}
\affiliation{Institute of Theoretical Physics, Technische
  Universit\"at Dresden, 01062 Dresden, Germany}
\author{Carsten Timm}
\email{carsten.timm@tu-dresden.de}
\affiliation{Institute of Theoretical Physics, Technische
  Universit\"at Dresden, 01062 Dresden, Germany}

\date{November 12, 2019}

\begin{abstract}
Noncentrosymmetric superconductors with line nodes are expected to possess topologically protected flat zero-energy bands of surface states, which can be described as Majorana modes. We here investigate their fate if residual interactions beyond BCS theory are included. For a minimal square-lattice model with a plaquette interaction, we find string-like integrals of motion that form Clifford algebras and lead to exact degeneracies. These degeneracies strongly depend on whether the numbers of sites in the \textit{x} and \textit{y} directions are even or odd, and are robust against disorder in the interactions. We show that the mapping of the Majorana model onto two decoupled spin compass models [Kamiya \textit{et al.}, Phys.\ Rev.\ B \textbf{98}, 161409 (2018)] and extra spectator degrees of freedom only works for open boundary conditions. The mapping shows that the three-leg and four-leg Majorana ladders are integrable, while systems of larger width are not. In addition, the mapping maximally reduces the effort for exact diagonalization, which is utilized to obtain the gap above the ground states. We find that this gap remains open if one dimension is kept constant and even, while the other is sent to infinity, at least if that dimension is odd. Moreover, we compare the topological properties of the interacting Majorana model to those of the toric-code model. The Majorana model has long-range entangled ground states that differ by $\mathbb{Z}_2$ fluxes through the system on a torus. The ground states exhibit string condensation similar to the toric code but the topological order is not robust. While the spectrum is gapped---due to spontaneous symmetry breaking inherited from the compass models---states with different values of the $\mathbb{Z}_2$ fluxes end up in the ground-state sector in the thermodynamic limit. Hence, the gap does not protect these fluxes against weak perturbations.
\end{abstract}

\maketitle

\section{Introduction}
\label{sec.intro}

From the discovery of superconductivity in 1911 to the very active and rapidly growing field of topological states of matter today, the field of condensed matter physics has seen the emergence of new paradigms. At the intersection of these fields, topological superconductors \cite{kiteavmajoraninwire, topsc1} exhibit fascinating properties of fundamental interest, such as the presence of Majorana quasiparticles in a condensed-matter system \cite{Beenakker2013, RevModPhys.87.137, Aguado2017}. Research is also driven by possible applications in fault-tolerant quantum computation~\cite{KITAEVquantumcomputing}.

Topological properties that emerge for effective single-electron models, in which interactions have notionally been treated at the mean-field level, are overall well understood. Topological invariants of single-electron Hamiltonians describing fully gapped insulators and superconductors have been obtained \cite{ryutenfold3D, tenfold3, tenfold4} based on the ten-fold-way classification by Zirnbauer and Altland \cite{Zir96, altlandtenfold_way}. However, unconventional superconductivity is often accompanied by zeros of the quasiparticle dispersion (relative to the Fermi energy), called gap nodes. The ten-fold-way classification for gapped systems has been extended to nodal systems \cite{PhysRevB.84.060504, doi:10.1143/JPSJ.81.011013, PhysRevLett.110.240404, matsuura_gapless, schnydersurface_nodal}, where topological invariants characterizing the nodes have been derived.

\begin{figure}[tb]
\includegraphics[width=0.6\columnwidth]{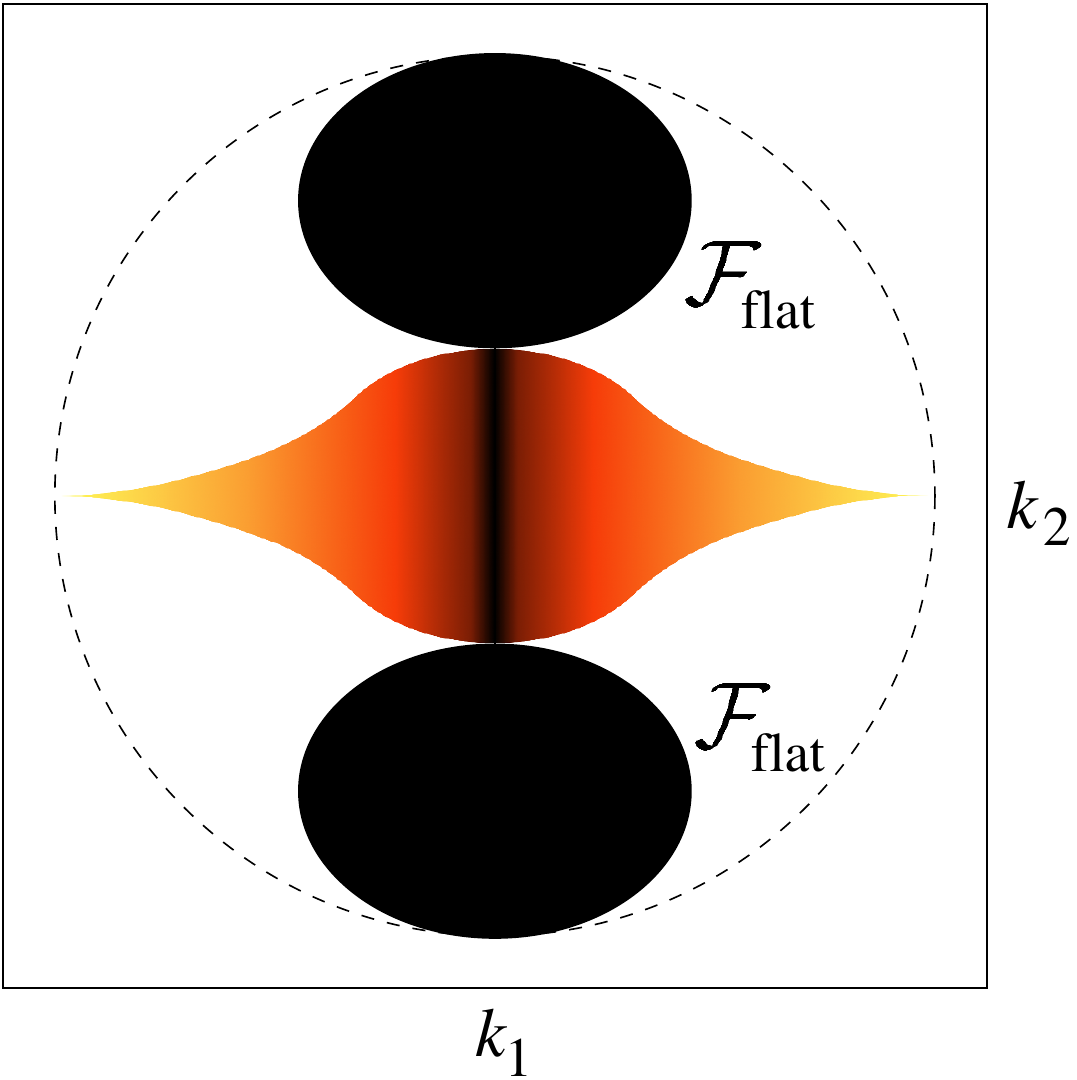}
\caption{\label{fig.C4v_surface}Cartoon of the dispersion of surface states at a $(101)$ or $(111)$ surface of a nodal NCS with point group $C_{4v}$ \cite{timmzerosurface_noncentro, timmtype_sur_state_nodal_noncentro}. Black areas denote flat zero-energy bands, color represents dispersive bands, and white means that there are no surface states at the corresponding momenta. The dashed line marks the projection of the edge of the Fermi sea.}
\end{figure}

Noncentrosymmetric superconductors (NCSs) are particularly interesting in this regard. The lack of inversion symmetry allows spin-orbit coupling that is odd in spin, which generically leads to Cooper pairs of mixed singlet-triplet character and, if the triplet pairing amplitude is sufficiently large, topologically protected line nodes \cite{PhysRevLett.87.037004, agterparallel, PhysRevB.84.060504, timmzerosurface_noncentro, timmtype_sur_state_nodal_noncentro, schnydersurface_nodal}. Promising candidates for such noncentrosymmetric systems are the heavy-fermion superconductors CePt$_3$Si and CeIrSi$_3$ \cite{CePt3Si,CeIrSi3} as well as the half-Heusler compounds YPtBi \cite{YPtBi, meinert, yptbisemi, Kimeaao4513, PhysRevB.96.094526} and LuPtBi~\cite{PhysRevB.87.184504}.
The line nodes are associated, by means of a bulk-boundary correspondence, with flat bands of surface states at zero energy (i.e., at the Fermi energy) \cite{flatbandNSC, flatbandYada, KHV11, HKV11, flatbandSato, PhysRevB.84.060504, timmzerosurface_noncentro, timmtype_sur_state_nodal_noncentro, schnydersurface_nodal, PhysRevB.96.094526}. Figure \ref{fig.C4v_surface} shows a cartoon of the resulting surface-state dispersion for a particular lattice symmetry and surface orientation.

The surface modes have the intriguing property of being their own antiparticles, i.e., they are Majorana modes. These modes were first predicted by Ettore Majorana in 1937 as elementary particles \cite{Majorana1937}, and are studied in a variety of contexts, from high-energy physics to quantum holography \cite{RevModPhys.87.137, SYKModel, PhysRevX.9.021043}. Besides the flat bands, other topological invariants can lead to the existence of additional arcs or points of zero-energy modes \cite{timmtype_sur_state_nodal_noncentro, schnydersurface_nodal, PhysRevB.96.094526}, which we do not consider in the following. One obvious question is whether the flat surface bands are stable. The system might reduce its free energy by shifting density of states away from the Fermi energy. Real-space BCS theory shows that this can indeed happen by spontaneous breaking of time-reversal symmetry in the surface region, at a temperature below the bulk transition~\cite{surfaceinstab}.

Another important question is what happens to the flat bands when residual interactions are included. More generally, topological properties of interacting systems are a very active field of research. Unlike for effectively noninteracting systems, no general classification scheme exists, at least not beyond one spatial dimension \cite{PhysRevB.83.075102, PhysRevB.83.075103, PhysRevB.85.075125, PhysRevB.96.165124}. However, significant insight has been gained by studying integrable models, in particular the toric-code model \cite{PhysRevB.40.7133, KITAEVquantumcomputing} and Kitaev's honeycomb model~\cite{KITAEVanyons}.

In this paper, we study the flat Majorana bands at the surface of NCSs in the presence of interactions. In Sec.\ \ref{sec.model1}, we discuss their theoretical modeling. In Sec.\ \ref{sec.model2}, we construct and analyze a minimal square-lattice model of interacting Majorana modes. We address its integrals of motion, degeneracies of states, and mapping to a spin compass model. The topological properties of the model are discussed in comparison to the toric-code model. A summary and conclusions are given in Sec.~\ref{sec.summary}.

\section{Topological surface states and their interactions}
\label{sec.model1}

As noted above, NCSs can have topologically protected line nodes in the bulk, associated with flat bands of surface states \cite{flatbandNSC, flatbandYada, flatbandSato, PhysRevB.84.060504, timmzerosurface_noncentro, timmtype_sur_state_nodal_noncentro, schnydersurface_nodal, PhysRevB.96.094526}. For time-reversal symmetric NCSs with line nodes, the winding number $W_{\mathcal{L}}(\mathbf{k}_{\|})$ is $\pm1$ if the momentum component parallel to the surface, $\mathbf{k}_{\|}$, lies within the projection of a single nodal line onto the two-dimensional surface Brillouin zone. We denote the corresponding subset of the surface Brillouin zone by $\mathcal{F}_\text{flat}$, corresponding to the black ellipses in Fig.\ \ref{fig.C4v_surface}, and the number of momenta $\mathbf{k}_{\|}$ within $\mathcal{F}_\text{flat}$ by $N_{\text{flat}}$. In the thermodynamic limit, $N_{\text{flat}}$ approaches infinity with $N_{\text{flat}}/N$ fixed, where $N$ is the total number of momenta in the surface Brillouin zone.

The diagonalization of the Bogoliubov-de Gennes Hamiltonian produces two zero-energy surface modes for each $\mathbf{k}_{\|}\in \mathcal{F}_\text{flat}$. Since the Bogoliubov-de Gennes-Nambu formalism double counts each fermionic degree of freedom, these correspond to a single physical mode per $\mathbf{k}_{\|}$. Denoting the corresponding quasiparticle annihilation operators by $\gamma_{\mathbf{k}_{\|}}$, we write the zero-energy modes in terms of Majorana operators
\begin{align}
\zeta_{\mathbf{k}_{\|}} &\equiv \gamma^\dagger_{\mathbf{-k}_{\|}}+\gamma^{\phantom{\dagger}}_{\mathbf{k}_{\|}} , \label{defmajo} \\
\tilde{\zeta}_{\mathbf{k}_{\|}}
  &\equiv i\, \big(\gamma^\dagger_{\mathbf{-k}_{\|}}-\gamma^{\phantom{\dagger}}_{\mathbf{k}_{\|}}\big) .
\end{align}
By an appropriate choice of phase factors of the $\gamma_{\mathbf{k}_{\|}}$, one can ensure that the two sets of Majorana modes are localized at the two surfaces of a NCS slab. The Majorana operators clearly satisfy $\zeta_{\mathbf{k}_{\|}}^\dagger = \zeta_{-\mathbf{k}_{\|}}^{\phantom{\dagger}}$ and $\tilde\zeta_{\mathbf{k}_{\|}}^\dagger = \tilde\zeta_{-\mathbf{k}_{\|}}^{\phantom{\dagger}}$.

We thus end up with $N_{\text{flat}}$ Majorana modes per surface, enumerated by $\mathbf{k}_{\|}\in \mathcal{F}_\text{flat}$. We are interested in the leading interactions beyond BCS theory. These can either be mediated by superconducting fluctuations about the saddle point \cite{Kle78,DCD99,TeD12,PCM15} or result from mechanisms not involved in the BCS decoupling. For example, there are magnetic dipolar interactions between the spin-polarized \cite{helical, edgecurrent, PhysRevB.96.094526} Majorana modes.

To construct an effective low-energy model, we choose to study only the flat zero-energy bands. Hence, the bilinear term in the Hamiltonian vanishes and the leading term is quartic. For a thick NCS slab, we may ignore interactions between the modes $\zeta_{\mathbf{k}_{\|}}$ and $\tilde\zeta_{\mathbf{k}_{\|}}$ localized at different surfaces. The Hamiltonian for one surface is then of the form~\cite{chiuinteract, PhysRevB.98.161409}
\begin{equation}
\label{Hint}
H = \frac{1}{4!}\sum_{ijkl}^{} g_{ijkl}\, \zeta_i\zeta_j\zeta_k\zeta_l,
\end{equation}
with the coupling tensor $g_{ijkl}$. The indices here label momenta $\mathbf{k}_{\|}\in\mathcal{F}_\text{flat}$. Note the similarity to the Sachdev-Ye-Kitaev (SYK) model \cite{SYK1, kitaevKITP1, kitaevKITP2, SYKADS}, where the coupling $g_{ijkl}$ is random. While the SYK model does not have any spatial structure and can thus be considered as zero dimensional, our model is two dimensional. 

The existence of zero-energy modes in a finite fraction $\mathcal{F}_\text{flat}$ of momentum space implies that one can construct wave packets localized at arbitrary positions in real space that are also eigenstates. Their minimal extension is inversely proportional to the typical diameter of $\mathcal{F}_\text{flat}$ in momentum space. The annihilation operators of maximally localized modes centered at positions $\mathbf{R}$ are given by
\begin{equation}
\Phi_{\mathbf{R}} = \frac{1}{\sqrt{N_{\text{flat}}}} \sum\limits_{\mathbf{k}_{\|}\in \mathcal{F}_\text{flat}}e^{i\mathbf{k}_{\|}\cdot\mathbf{R}}\, \zeta_{\mathbf{k}_{\|}}.
\end{equation}
From $\zeta_\mathbf{k}^\dagger = \zeta_{-\mathbf{k}}$ one obtains the Majorana property $\Phi_\mathbf{R}^\dagger = \Phi_\mathbf{R}$. However, since the flat band does not exist in the whole two-dimensional Brillouin zone, the set of $N$ modes described by $\Phi_{\mathbf{R}}$ centered at all positions $\mathbf{R}$ is overcomplete---there can only be $N_\text{flat}$ independent modes. This is also shown by the nontrivial anticommutation relation
\begin{equation}
\{ \Phi_\mathbf{R}, \Phi_{\mathbf{R}'} \} = \frac{2}{N_\text{flat}} \sum\limits_{\mathbf{k}_{\|}\in \mathcal{F}_\text{flat}} e^{i\mathbf{k}_{\|}\cdot(\mathbf{R}-\mathbf{R}')}\: \mathbbm{1} .
\end{equation}
Put differently, the modes have nonvanishing overlap integrals~\cite{Loewdin}
\begin{equation}
S_{\mathbf{R}\mathbf{R}'} = \frac{1}{N_{\text{flat}}}\sum\limits_{\mathbf{k}_{\|}\in \mathcal{F}_\text{flat}}e^{i \mathbf{k}_{\|}\cdot(\mathbf{R}-\mathbf{R}')}-\delta_{\mathbf{R}\mathbf{R}'}.
\label{eq.SRRp}
\end{equation}
Such nonvanishing overlaps can also be interpreted in terms of quantum (noncommutative) geometry \cite{doi:10.1063/1.5046122}. To construct a model in real space, it is necessary to first choose $N_\text{flat}$ real-space points $\mathbf{R}$ and then construct an orthonormal basis out of the wave packets localized at these points. We emphasize that we are free to choose the points $\mathbf{R}$ as long as we ensure to have $N_{\text{flat}}$ Majorana modes with the correct density. We choose a square lattice since it will allow for a natural approximation for the interaction in the next step. The lattice constant $a$ must then satisfy $N_\text{flat}/N = s_\text{uc}/a^2$, where $s_\text{uc}$ is the area of the two-dimensional surface unit cell of the microscopic lattice.

Following L{\"o}wdin \cite{Loewdin}, we construct an orthonormal basis using the overlap matrix $S$ with the components $S_{\mathbf{R}\mathbf{R}'}$. Note that $S$ is real and symmetric because the region $\mathcal{F}_\text{flat}$ is symmetric with respect to the center of the two-dimensional Brillouin zone. The sequence
\begin{equation}
\vec\zeta \equiv \big( \zeta_{\mathbf{R}_1}, \zeta_{\mathbf{R}_2}, \ldots, \zeta_{\mathbf{R}_{N_\text{flat}}}\big)
\end{equation}
of operators $\zeta_\mathbf{R}$ describing orthonormal states is then related to the sequence
\begin{equation}
\vec{\Phi} \equiv \big(\Phi_{\mathbf{R}_1}, \Phi_{\mathbf{R}_2}, \ldots, \Phi_{\mathbf{R}_{N_{\text{flat}}}}\big)
\end{equation}
of operators $\Phi_\mathbf{R}$ describing independent but not orthonormal states by
\begin{equation}
\vec{\Phi} = \vec{\zeta}\: (\mathbbm{1}+S)^{1/2} ,
\label{orthonorms}
\end{equation}
and conversely
\begin{equation}
\vec{\zeta} = \vec{\Phi}\: (\mathbbm{1}+S)^{-1/2}.
\label{orthonorms2}
\end{equation}
The matrix root is understood in terms of the usual power series.
Since $S$ is real the property $\zeta_\mathbf{R}^\dagger = \zeta_\mathbf{R}$ is retained. Furthermore, by inserting Eq.\ (\ref{orthonorms2}), we find the canonical anticommutation relation $\{\zeta_{\mathbf{R}},\zeta_{\mathbf{R}'}\} = 2\delta_{\mathbf{R},\mathbf{R}'}$. While the modes described by $\Phi_{\mathbf{R}}$ are maximally localized, the localization of the transformed modes depends on the matrix $S$. Roughly speaking, the L{\"o}wdin method yields the orthonormal set that is most similar to the original functions \cite{Loewdin}. Hence, the main weight is still located at $\mathbf{R}$ but the state is smeared out over all lattice sites, weighted by powers of the matrix $S$, which depends on the material-specific region $\mathcal{F}_\text{flat}$.

For illustration, we present the overlap matrix for the example of $\mathcal{F}_\text{flat}$ consisting of two elliptical regions as in Fig.\ \ref{fig.C4v_surface}. For two ellipses centered at $\pm\mathbf{Q}=\pm(Q_1,Q_2)$ with semi-axes $c_1$ and $c_2$, we obtain
\begin{align}
S_{\mathbf{R}\mathbf{R}'} &= 2a \cos\left( \left(\frac{Q_1}{c_1}, \frac{Q_2}{c_2}\right)
  \cdot \frac{\mathbf{R}-\mathbf{R}'}{a} \right) \nonumber \\
&\quad{}\times \frac{J_1(|\mathbf{R}-\mathbf{R}'|/a)}{|\mathbf{R}-\mathbf{R}'|}
    - \delta_{\mathbf{R}\mathbf{R}'} ,
\end{align}
where $J_1(x)$ is a Bessel function. Since $\mathcal{F}_\text{flat}$ covers a fraction of $N_\text{flat}/N$ of the Brillouin zone, the semi-axes satisfy $c_1 c_2 = 2\pi/a^2$. Note that the envelope of $S_{\mathbf{R}\mathbf{R}'}$ decays as a power law of the distance $|\mathbf{R}-\mathbf{R}'|$.

The Hamiltonian in terms of $\vec{\Phi}$ can be obtained by Fourier transforming the momentum-space Hamiltonian (\ref{Hint}). Writing the result as
\begin{equation}
H = \frac{1}{4!} \sum\limits_{ijkl} \tilde{g}_{\mathbf{R}_i\mathbf{R}_j\mathbf{R}_k\mathbf{R}_l}\,
  \Phi_{\mathbf{R}_1} \Phi_{\mathbf{R}_2} \Phi_{\mathbf{R}_3} \Phi_{\mathbf{R}_4}
\end{equation}
and substituting Eq.\ (\ref{orthonorms}), we find
\begin{align}
\label{bilin}
H &= \frac{1}{4!} \sum\limits_{ijkl} \tilde{g}_{\mathbf{R}_i\mathbf{R}_j\mathbf{R}_k\mathbf{R}_l}\,
  (\vec{\zeta}\,(\mathbbm{1}+S)^{1/2})_{\mathbf{R}_i}
  (\vec{\zeta}\,(\mathbbm{1}+S)^{1/2})_{\mathbf{R}_j} \nonumber \\
&\quad{}\times (\vec{\zeta}\,(\mathbbm{1}+S)^{1/2})_{\mathbf{R}_k}
  (\vec{\zeta}\,(\mathbbm{1}+S)^{1/2})_{\mathbf{R}_l}.
\end{align}
We can now redefine the coupling according to
\begin{align}
g_{\mathbf{R}_i\mathbf{R}_j\mathbf{R}_k\mathbf{R}_l} &= \sum\limits_{mnop} \tilde{g}_{\mathbf{R}_m\mathbf{R}_n\mathbf{R}_o\mathbf{R}_p}
  (\mathbbm{1}+S)^{1/2}_{\mathbf{R}_i\mathbf{R}_m}
  (\mathbbm{1}+S)^{1/2}_{\mathbf{R}_j\mathbf{R}_n} \nonumber \\
&\quad{}\times (\mathbbm{1}+S)^{1/2}_{\mathbf{R}_k\mathbf{R}_o}
  (\mathbbm{1}+S)^{1/2}_{\mathbf{R}_l\mathbf{R}_p} .
\end{align}
Identifying the subscript $\mathbf{R}_i$ with the index $i$, we can write the Hamiltonian as $H = {1}/{4!}\, \sum_{ijkl} g_{ijkl}\, \zeta_i \zeta_j \zeta_k \zeta_l$, which is formally identical to Eq.\ (\ref{Hint}) but now pertains to real space. Note that the real-space Hamiltonian is equivalent to the original one in momentum space for \emph{any} choice of wave-packet centers $\mathbf{R}$ with the correct density.

We have here obtained a new platform for interacting Majorana modes in two dimensions. Previously, such models were derived for Majorana modes bound to vortices in two-dimensional topological superconductors \cite{Volovik1999, PhysRevLett.100.096407, chiuinteract, PhysRevB.98.161409}. In our case, the absence of a bilinear term is due to the topological winding number of bulk line nodes and, unlike for the realization in a vortex lattice, does not require fine tuning of the chemical potential. The model with a bilinear term has been studied by Affleck \textit{et al.}\ \cite{ARP17} using mean-field and renormalization-group methods.

\section{Interacting Majorana modes on a square lattice}
\label{sec.model2}

As shown in the previous section, the coupling tensor $g_{ijkl}$ in real space can be obtained from the one in momentum space by a Fourier transformation followed by orthonormalization. The interaction is expected to decay like a power law with separation, due to the power-law decay of the orthonormalized states. General properties of the coupling $g_{ijkl}$ are dictated by fundamental requirements: It is real due to hermiticity and can be chosen to be completely antisymmetric since the $\zeta_i$ anticommute. In particular, $g_{ijkl}$ is zero if two indices are equal.

Symmetries constrain $g_{ijkl}$ further \footnote{Time-reversal symmetry is trivial for our model. Since the positions are invariant under time reversal and the Majorana modes are nondegenerate the only effect of time reversal on $\zeta_i$ could be a sign change. A more general phase factor would be incompatible with the Majorana property. Translation symmetry of the lattice then requires all Majorana modes to be either even or odd under time reversal. The sign drops out of the Hamiltonian and in particular does not lead to a constraint on $g_{ijkl}$.}. If the system is invariant under the transformation $\zeta_i \to \sum_j O_{ij} \zeta_j$ with an orthogonal matrix $O$ then the couplings must satisfy
\begin{equation}
\sum_{mnop} g_{mnop}\, O_{mi} O_{nj} O_{ok} O_{pl} = g_{ijkl} .
\end{equation}
The transformation matrix must be orthogonal to preserve the Majorana property. In the following, we construct a minimal model in real space and study its ground state, order, and low-energy excitations.

\subsection{Minimal model on the square lattice}

In order to construct a minimal model, we truncate the interaction after the most localized term, based on the expectation that the interaction decays with separation. Here, the choice of lattice in Sec.\ \ref{sec.model1} becomes important---we should choose the lattice in such a way that the truncation is a reasonable approximation. This is the case for the square lattice, which has a natural most strongly localized contribution, namely the plaquette terms of four Majorana modes localized at the corners of elementary squares. This means that we take $g_{ijkl} = g$ if $\mathbf{R}_i$, $\mathbf{R}_j$, $\mathbf{R}_k$, and $\mathbf{R}_l$ belong to the same plaquette and zero otherwise.

We note that long-range interactions may have interesting consequences, which we leave for future work. To fix the order of the anticommuting Majorana operators, we use plaquette operators $p_i \equiv \zeta_{\mathbf{R}_i} \zeta_{\mathbf{R}_i+\mathbf{a}_x} \zeta_{\mathbf{R}_i+\mathbf{a}_x+\mathbf{a}_y} \zeta_{\mathbf{R}_i+\mathbf{a}_y}$, where $\mathbf{a}_{x,y}$ are the lattice basis vectors. The Hamiltonian is then given by
\begin{equation}
H = g \sum\limits_{i} p_i .
\label{Hint2}
\end{equation}
In the following, we consider lattices with $L_x\times L_y = N_\text{flat}$ sites and periodic or open boundary conditions in either direction. The $N_\text{flat}$ Majorana modes $\zeta_i$ can be re-expressed in terms of $N_\text{flat}/2$ complex fermions. The Hilbert-space dimension is thus $2^{N_\text{flat}/2} = 2^{L_xL_y/2}$. This is evidently impossible if both $L_x$ and $L_y$ are odd \footnote{An odd number $L_xL_y$ of Majorana modes is impossible for a model where $\mathcal{F}_\text{flat}$ consists of an even number of regions as in Fig.\ \ref{fig.C4v_surface}. However, if we include the arc of zero-energy states, which contains a mode at $\mathbf{k}=0$, their number is odd. Whether a zero mode at $\mathbf{k}=0$ exists, depends on symmetry \cite{timmzerosurface_noncentro, timmtype_sur_state_nodal_noncentro}.}, which signifies that the two surfaces cannot be treated separately in that case. However, considering Eq.\ (\ref{Hint2}) without regard to its origin, we can treat the odd times odd case by introducing an additional ``spectator'' Majorana mode $\zeta_*$ that does not appear in the Hamiltonian but makes the number of Majorana modes even.

Equation (\ref{Hint2}) is closely related to the toric-code model \cite{PhysRevB.40.7133, KITAEVquantumcomputing}. The toric-code model is usually defined in terms of spin operators located at the edges of a square lattice. However, these spins also form the vertices of a rotated square lattice. In this representation, the model is described by the Hamiltonian
\begin{align}
H_\text{toric code} &= g \sum\limits_{i\in A} \sigma^x_{\mathbf{R}_i} \sigma^x_{\mathbf{R}_i+\mathbf{a}_x}
  \sigma^x_{\mathbf{R}_i+\mathbf{a}_x+\mathbf{a}_y} \sigma^x_{\mathbf{R}_i+\mathbf{a}_y} \nonumber \\
&\quad{}+ g \sum\limits_{i\in B} \sigma^z_{\mathbf{R}_i} \sigma^z_{\mathbf{R}_i+\mathbf{a}_x}
  \sigma^z_{\mathbf{R}_i+\mathbf{a}_x+\mathbf{a}_y} \sigma^z_{\mathbf{R}_i+\mathbf{a}_y} ,
\end{align}
where the sums are over sites of the two checkerboard sublattices $A$ and $B$ and $\sigma^{x,z}_{\mathbf{R}}$ are spin operators (suppressing factors $\hbar/2$). The toric-code model is integrable since all plaquette terms commute \cite{KITAEVquantumcomputing}. Its spectrum is discrete and, in particular, has an energy gap $2|g|$ above the ground state.

In contrast, in our model, two plaquette operators $p_i$ and $p_j$ commute if they share zero or two lattice sites but anticommute if they share only a single one, i.e, a corner. Because of the noncommutativity of the plaquettes, the model is not integrable \cite{chiuinteract}. This distinction leads to different properties, as we shall see.

Chiu \textit{et al.}\ \cite{chiuinteract} have studied a range of models of interacting Majorana modes, including the present one. Based on exact diagonalization for small systems, they have concluded that the model with uniform plaquette couplings is gapless \cite{chiuinteract}. However, this work has been superseded by Kamiya \textit{et al.}\ \cite{PhysRevB.98.161409}, who study the same model by means of Jordan-Wigner mappings to spin models and quantum Monte Carlo simulations. The authors find clear evidence for a finite-temperature phase transition towards a gapped low-temperature phase with stripe order \cite{PhysRevB.98.161409}. We return to this point below.


\subsection{Symmetries and invariants}
\label{sub.invar}

Symmetries can be exploited to simplify the solution and to better understand the system. For this, it is useful to note that any product $P = \zeta_1 \cdots \zeta_n$ of Majorana operators is unitary: $P^\dagger P = \zeta_n \cdots \zeta_1 \zeta_1 \cdots \zeta_n = \mathbbm{1}$. Moreover, such a product is hermitian (antihermitian) if $n \bmod 4 \in \{0,1\}$ ($\{2,3\}$) since it takes an even (odd) number of pair exchanges to transform $P^\dagger$ into $P$.

In case of open boundary conditions, we can construct a unitary operator $\mathcal{C}$ that anticommutes with the Hamiltonian $H$ by forming the product of one Majorana operator from each plaquette. We can think of $\mathcal{C}$ as a charge-conjugation operator. Its existence guarantees that the spectrum is symmetric. For periodic boundary conditions, the charge-conjugation operator can only be constructed for even times even numbers of lattice sites.

The model with periodic or open boundary conditions has a large number of integrals of motion, among them the products of all Majorana operators in row $y$ or column $x$ of the lattice. We denote these products as ``row operators''
\begin{equation}
R_y = \left\{\begin{array}{cl}
  1 & \mbox{for $L_x \bmod 4 \in \{0,1\}$} \\
  i & \mbox{for $L_x \bmod 4 \in \{2,3\}$}
  \end{array}\right\} \times \prod\limits_{x=1}^{L_x} \zeta_{x,y}
\end{equation}
and ``column operators''
\begin{equation}
C_x = \left\{\begin{array}{cl}
  1 & \mbox{for $L_y \bmod 4 \in \{0,1\}$} \\
  i & \mbox{for $L_y \bmod 4 \in \{2,3\}$}
  \end{array}\right\} \times \prod\limits_{y=1}^{L_y} \zeta_{x,y} ,
\end{equation}
respectively \cite{chiuinteract}, where the conditional factors ensure hermiticity and guarantee that the operators square to $+\mathbbm{1}$. These integrals of motion realize one-dimensional gauge-like symmetries in the sense of Batista and Nussinov \cite{PhysRevB.72.045137}. The important consequence is that the existence of local order parameters is governed by a one-dimensional effective Hamiltonian \cite{PhysRevB.72.045137}. Hence, the model cannot have a nonvanishing local order parameter at temperatures $T>0$, except if the order parameter commutes with the row and column operators.

\begin{table*}[tb]
\renewcommand{\arraystretch}{1.2}
\begin{tabular}{|c|c|c|}
\hline even $\times$ even & even $\times$ odd & odd $\times$ odd \\
\hline\hline
$[R_y,R_{y'}]=0$ & $[R_y,R_{y'}]=0$ & $[R_y,C_x]=0$ \\
$[C_x,C_{x'}]=0 $ & $[\Gamma_{xy},\Gamma_{xy'}]=0$ & $[R_y,\Gamma_{xy}]=0$ \\
$[\Gamma_{xy},\Gamma_{x'y'}]=0$ & $ [C_x,\Gamma_{x'y'}]=0$ for $x\ne x'$ & $[C_x,\Gamma_{xy}]=0$ \\
 & & $[\Gamma_{xy},\Gamma_{x'y'}]=0$ for $x\neq x'$ and $y\neq y'$ \\
\hline $\{R_y,C_x\}=0$ & $\{C_x,C_{x'}\}=2\delta_{xx'}\mathbbm{1}$ & $\{R_y,R_{y'}\}=2\delta_{yy'}\mathbbm{1}$ \\
$\{R_y,\Gamma_{x'y'}\}=0$ & $\{\Gamma_{xy},\Gamma_{x'y'}\}=0$ for $x\ne x'$ & $\{C_x,C_{x'}\}=2\delta_{xx'}\mathbbm{1}$ \\
$\{C_x,\Gamma_{x'y'}\}=0$ & $\{R_y,C_x\}=0$ & $\{\Gamma_{xy},\Gamma_{xy'}\}=0$ for $y\neq y'$ \\
 & $\{R_y,\Gamma_{x'y'}\}=0$ & $\{\Gamma_{xy},\Gamma_{x'y}\}=0$ for $x\neq x'$ \\
 & $\{C_x,\Gamma_{xy}\}=0$ & $\{R_y,\Gamma_{x'y'}\}=0$ for $y\neq y'$ \\
 & & $\{C_x,\Gamma_{x'y'}\}=0$ for $x\neq x'$ \\
\hline
\end{tabular} 
\caption{Commutation relations of the row operators $R_y$, column operators $C_x$, and cross operators $\Gamma_{xy} = iC_xR_y$. For the case of odd times even, rows and columns have to be interchanged.}
\label{Commrules}
\end{table*}

We first discuss the even times even lattice. In this case, the row and column operators all commute among themselves but anticommute between different types, see table \ref{Commrules}. We further define the ``cross operators'' as the products $\Gamma_{xy} = iC_xR_y$. They contain all Majorana operators in one row and one column, except for their crossing point. The cross operators also commute among themselves but anticommute with the row and column operators. Additionally, all row, column, and cross operators square to~$+\mathbbm{1}$. This means that for arbitrary but fixed $x$ and $y$, the three operators $R_y$, $C_x$, and $\Gamma_{xy}$ satisfy the algebra of the Pauli matrices $\sigma^1$, $\sigma^2$, and $\sigma^3$.

Hence, the model has $L_xL_y = N_\text{flat}$ integrals of motion $\Gamma_{xy}$ that commute among themselves and have eigenvalues $\pm 1$. Nevertheless, the model is not integrable since these invariants are not independent. Rather, the cross operators are subject to the constraints $\Gamma_{xy}\Gamma_{x'y'} = \Gamma_{xy'}\Gamma_{x'y}$. Consequently, only $L_x+L_y-2$ of the $\Gamma_{xy}$ are independent since specifying the invariants in one row and one column fixes all of them. This involves $L_x+L_y-1$ cross operator but only $L_x+L_y-2$ of them are independent since the product of all $\Gamma_{xy}$ for a single row equals the product for a single column, except possibly for a sign.

\begin{figure}
\includegraphics[width=0.65\columnwidth,clip]{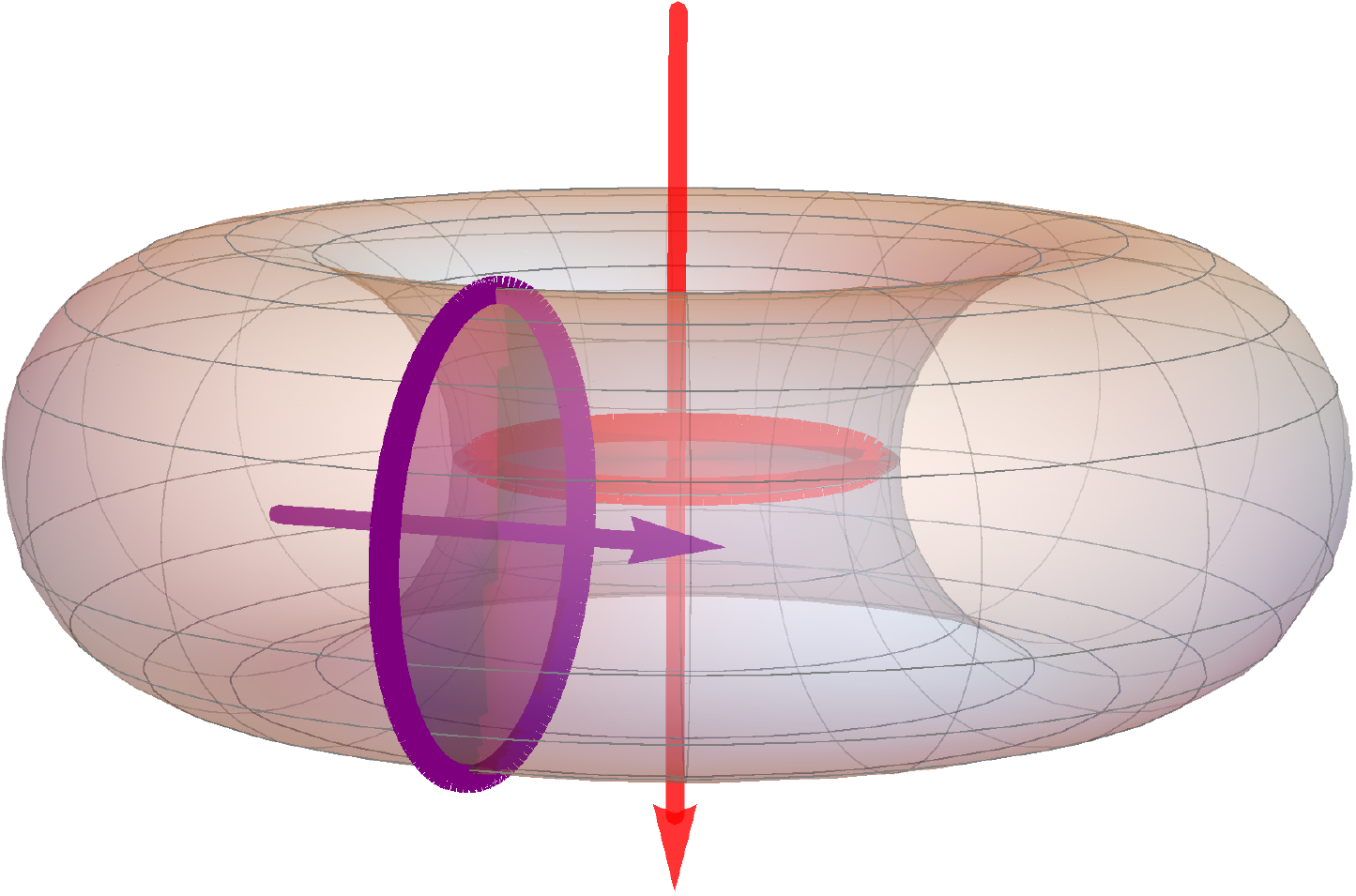}
\caption{Visualization of a row operator $R_y$ (red line) and a column operator $C_x$ (purple line) as loops on a torus, for periodic boundary conditions. We call their eigenvalues ``fluxes,'' in analogy to the $\mathbb{Z}_2$ fluxes in the toric-code model \cite{wenquantumorder, booklinkedcluster}. The fluxes are illustrated by the arrows.}
\label{fig.torusflux}
\end{figure}

For periodic boundary conditions, the row and column operators define loops on a torus, as sketched in Fig.\ \ref{fig.torusflux}. Since these operators have eigenvalues $\pm 1$, we can think of them as $\mathbb{Z}_2$ \emph{fluxes} penetrating the torus, in analogy to the toric-code model~\cite{wenquantumorder, booklinkedcluster}.

For the even times odd lattice, the column operators $C_x$ anticommute pairwise since they are products of odd numbers of Majorana operators, see table \ref{Commrules}. For $L_y>3$, this leads to a larger number of mutually anticommuting operators that commute with the Hamiltonian than for the even times even lattice. The consequences for the degeneracy of states are discussed below. Nevertheless, the triple $R_y$, $C_x$, and $\Gamma_{xy}$ still satisfies the algebra of the Pauli matrices. The odd times even case is of course analogous.

For the odd times odd lattice, both the row and the column operators anticommute among themselves and the other relations become more complicated, see table \ref{Commrules}. There is no triple of operators that realize the Pauli algebra. However, including the spectator Majorana mode $\zeta_*$, we can find such triples, for example $R_y$, $\zeta_*$, and $i\zeta_*R_y$.

\subsection{Degeneracies}

In this section, we consider degeneracies resulting from the integrals of motion. They turn out to depend dramatically on whether $L_x$ and $L_y$ are even or odd. The degeneracies are topological in the sense that they are preserved under random perturbations of the plaquette couplings $g_{ijkl}$. For random couplings, the degeneracy is the same for all energy levels. For uniform $g_{ijkl}$, as considered in the previous sections, lattice symmetries lead to additional degeneracies. Our results also hold for open or mixed boundary conditions.

Our analysis is based on the theory of Clifford algebras \cite{Clifford1}. A number $n=1,2,\ldots$ of operators that square to $+\mathbbm{1}$ and anticommute with each other generate the Clifford algebra $\mathrm{C}\ell_n(\mathbb{C})$ on the vector space $\mathbb{C}^n$ with the standard bilinear form. If $n=2m$ is even, $\mathrm{C}\ell_{2m}(\mathbb{C})$ is isomorphic to the algebra of complex matrices of dimension $2^m$. On the other hand, if $n=2m+1$ is odd, $\mathrm{C}\ell_{2m+1}(\mathbb{C})$ is isomorphic to the direct sum of two copies of the algebra of complex matrices of dimension $2^m$. In other words, for even (odd) $n$, the Clifford algebra has one irreducible matrix representation (two irreducible representations) of dimension $2^{\lfloor n/2\rfloor}$, where $\lfloor x\rfloor$ is the largest integer not greater than $x$. If the $n$ anticommuting operators commute with the Hamiltonian, the degeneracy of all eigenenergies contains a factor of $2^{\lfloor n/2\rfloor}$.

For even times even sites, we have found three pairwise anticommuting integrals of motion, namely $R_y$, $C_x$, and $\Gamma_{xy}$ for arbitrary but fixed $x$, $y$. There is no additional row, column, or cross operator that anticommutes with all three and thus we have $n=3$, leading to a degeneracy of $2^{\lfloor 3/2\rfloor}=2$.

For even times odd sites, the column operators $C_x$ anticommute pairwise, see table \ref{Commrules}. There are $L_x$ of them, which is an even number. One can find one additional operator, namely any $R_y$ or any $\Gamma_{xy}$, that anticommutes with all $C_x$, which does not increase the degeneracy. Hence, the degeneracy is $2^{L_x/2}$.

For odd times odd sites, the row operators and the column operators anticommute among themselves but not with each other, which suggests a degeneracy of $\max(2^{\lfloor L_x/2\rfloor}, 2^{\lfloor L_y/2\rfloor})$. However, the actual degeneracies are larger: As discussed above, the model requires the introduction of a spectator Majorana mode $\zeta_*$ to obtain an even number of modes. We then find
\begin{align}
\{R_y,R_{y'}\} &= 2\delta_{yy'}\mathbbm{1} , \\
\{\zeta_* C_x, \zeta_* C_{x'}\} &= 2\delta_{xx'}\mathbbm{1} , \\
\{R_y, \zeta_* C_x\} &= 0 .
\end{align}
Thus the $L_x+L_y$ (which is even) integrals of motion $R_y$ and $\zeta_* C_x$ anticommute pairwise, leading to a larger degeneracy of $2^{(L_x+L_y)/2}$. All of these results are corroborated by exact diagonalization for small systems with random plaquette couplings $g_{ijkl}$, indicating that we have indeed found the largest number of anticommuting integrals of motion. The degeneracies are summarized in table~\ref{tab.degeneracy}. We note that since the degeneracies are independent of open vs.\ periodic boundary conditions they cannot be attributed to decoupled modes localized at the edges.

\begin{table}[htb]
\begin{tabular}{|c|c|} \hline
$L_x \times L_y$ & degeneracy \\ \hline\hline
even $\times$ even & $2$ \\
even $\times$ odd & $2^{L_x/2}$ \\
odd $\times$ even & $2^{L_y/2}$ \\
odd $\times$ odd & $2^{(L_x+L_y)/2}$ \\ \hline
\end{tabular} 
\caption{Degeneracies of states for all combination of even and odd sizes $L_x$ and $L_y$. Lattice symmetries may lead to additional degeneracies, which multiply the ones given here. The degeneracy of the ground state is always the one shown.}
\label{tab.degeneracy}
\end{table}

\subsection{Mapping to compass models}
\label{sub.mapping}

\begin{figure}[thb]
\includegraphics[scale=0.6]{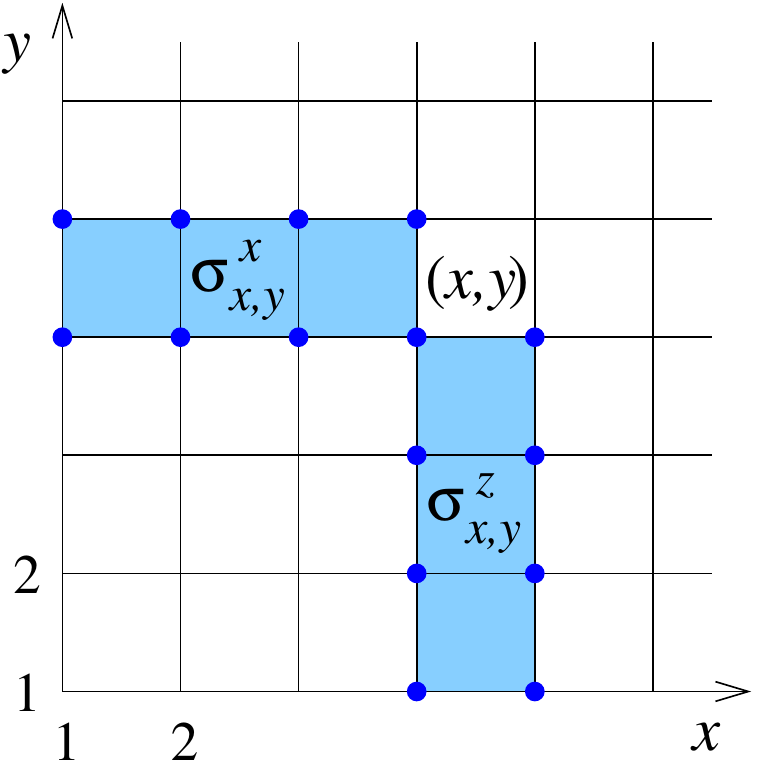}
\caption{Illustration of the locations of Majorana modes appearing in the mapped spin operators $\sigma^x_{x,y}$ and $\sigma^z_{x,y}$ at plaquette $(x,y)$. The sites of Majorana modes are indicated by blue circles. One of them appears in both spin operators, ensuring that they anticommute.}
\label{fig.compass_map1}
\end{figure}

Kamiya \textit{et al.}\ \cite{PhysRevB.98.161409} present a three-step mapping of the interacting Majorana model onto two decoupled quantum compass models by way of two intermediate spin models. In the following, we describe a direct mapping of the Majorana model with open boundary conditions to two decoupled compass models. We restrict ourselves to $L_x,L_y>2$ since the two-leg ladder is nongeneric and trivially integrable \cite{chiuinteract}. In this mapping, a quantum spin $\mbox{\boldmath$\sigma$}_{x,y}$ of length $1/2$ is associated with half of the plaquettes of the original model. We enumerate the plaquettes in such a way that the plaquette at $(x,y)$ involves the product $\zeta_{x,y} \zeta_{x+1,y} \zeta_{x+1,y+1} \zeta_{x,y+1}$. The mapping must be such that different spin components at the same site anticommute, whereas spin operators at different sites always commute. In the first step, we set
\begin{align}
\sigma^x_{x,y} &= \left\{\begin{array}{cl}
    1 & \mbox{for $x$ even} \\
    i & \mbox{for $x$ odd}
  \end{array} \right\} \times \prod\limits_{x'=1}^{x} \zeta_{x',y} \zeta_{x',y+1} ,
\label{eq.cm.sigmax} \\
\sigma^z_{x,y} &= i^y \prod\limits_{y'=1}^{y} \zeta_{x,y'} \zeta_{x+1,y'} .
\label{eq.cm.sigmaz}
\end{align}
The locations of Majorana modes appearing in the definitions are illustrated in Fig.\ \ref{fig.compass_map1}. The numerical factors ensure that the spin operators square to $+\mathbbm{1}$ and that the plaquette terms do not contain extra signs:
\begin{align}
\sigma^x_{x,y} \sigma^x_{x+2,y}
  &= \zeta_{x+1,y} \zeta_{x+2,y} \zeta_{x+2,y+1} \zeta_{x+1,y+1}
\label{eq.cmmapx} , \\
\sigma^z_{x,y} \sigma^z_{x,y+2}
  &= \zeta_{x,y+1} \zeta_{x+1,y+1} \zeta_{x+1,y+2} \zeta_{x,y+2} .
\label{eq.cmmapz}
\end{align}
These equations imply
\begin{equation}
\zeta_{x,y} \zeta_{x+1,y} \zeta_{x+1,y+1} \zeta_{x,y+1}
  = \sigma^x_{x-1,y} \sigma^x_{x+1,y}
  = \sigma^z_{x,y-1} \sigma^z_{x,y+1} .
\label{eq.cmmapp}
\end{equation}
Note that the plaquette terms correspond to \emph{edges} connecting two spins that are two units apart. It is easy to see that for the first row or column the plaquette term is represented by one spin operator alone, e.g., $\zeta_{1,y} \zeta_{2,y} \zeta_{2,y+1} \zeta_{1,y+1} = \sigma^x_{2,y}$. In order to use Eq.\ (\ref{eq.cmmapp}) for all plaquettes, we extend the definitions in Eqs.\ (\ref{eq.cm.sigmax}) and (\ref{eq.cm.sigmaz}) such that for $x\le 0$ or $y\le 0$ the product is understood as equaling~$\mathbbm{1}$.

The full set of operators in Eqs.\ (\ref{eq.cm.sigmax}) and (\ref{eq.cm.sigmaz}) does not satisfy the correct algebra of a spin model. Rather, two $\sigma^x$ for adjacent rows $y$ and $y+1$ and odd $x$ as well as two $\sigma^z$ for adjacent columns $x$ and $x+1$ and odd $y$ anticommute since they have an odd number of Majorana modes in common. To avoid the first problem, we only use spin operators in every second row, and to avoid the second, we specifically take even-numbered rows (even $y$) \footnote{Of course one can interchange rows and columns in everything that follows.}. Then the only anticommuting combinations are $\sigma^x_{x,y}$ and $\sigma^z_{x,y}$ since only such pairs have an odd number of Majorana modes, namely a single one, in common, see Fig.\ \ref{fig.compass_map1}. The restricted set of spin operators still allows to express all plaquette terms by
using Eqs.\ (\ref{eq.cmmapx}) and (\ref{eq.cmmapz}) for alternating rows.

\begin{figure}[bt]
\includegraphics[scale=0.6]{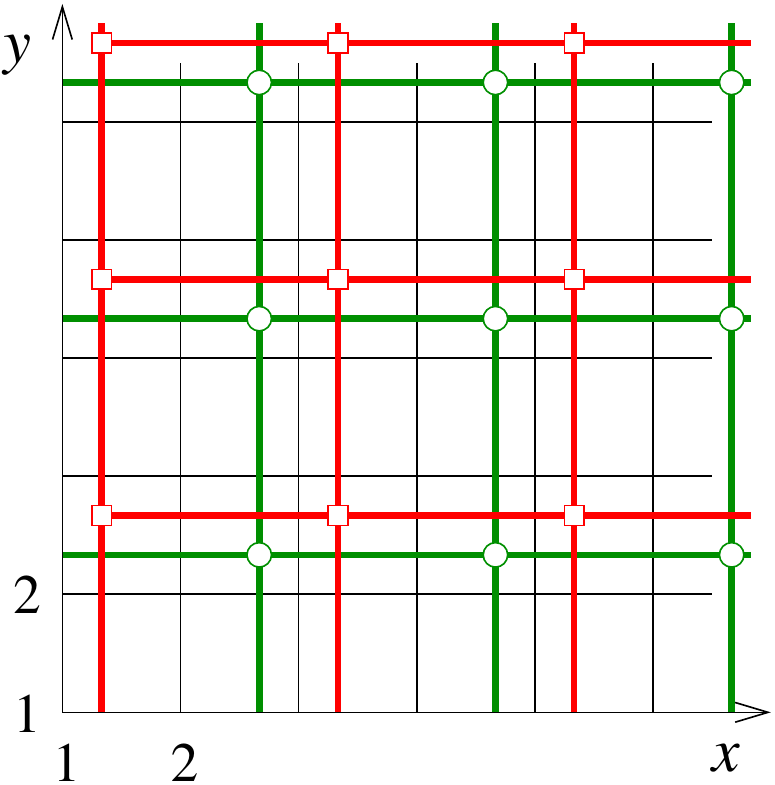}
\caption{Illustration of the compass models resulting from the mapping of the interacting Majorana model. The locations of the spins are denoted by red squares (subsystem 1) and green circles (subsystem 2), which are displaced diagonally for clarity. Bonds are denoted by heavy red and green lines connecting the spins. Note that there is a single bond in every plaquette of the original lattice.}
\label{fig.compass_map2}
\end{figure}

The Hamiltonian then reads
\begin{equation}
H = g \sum_x \sum_{y\:\text{even}} \left( \sigma^x_{x,y} \sigma^x_{x+2,y}
  + \sigma^z_{x,y} \sigma^z_{x,y+2} \right) .
\label{eq.H.compass.2}
\end{equation}
The terms appearing here are illustrated in Fig.\ \ref{fig.compass_map2}. As expected \cite{PhysRevB.98.161409}, there is no coupling between terms involving spins with even and odd $x$ coordinates so that the model decomposes into two decoupled compass models. In the following, we will denote the compass subsystem involving spins at odd (even) $x$ as subsystem 1~(2).

\begin{table}
\begin{tabular}{|c|c|c|} \hline
  & subsystem 1 & subsystem 2 \\
\raisebox{1.2ex}[1.2ex]{$L_x \times L_y$} & bottom top left right & bottom top left right \\ \hline\hline
even $\times$ even & \quad \yes \hfill \yes \hfill \no \hfill \no \mbox{\quad} &
  \quad \yes \hfill \yes \hfill \yes \hfill \yes \mbox{\quad} \\ \hline
even $\times$ odd & \quad \yes \hfill \no \hfill \no \hfill \no \mbox{\quad} &
  \quad \yes \hfill \no \hfill \yes \hfill \yes \mbox{\quad} \\ \hline
odd $\times$ even & \quad \yes \hfill \yes \hfill \no \hfill \yes \mbox{\quad} &
  \quad \yes \hfill \yes \hfill \yes \hfill \no \mbox{\quad} \\ \hline
odd $\times$ odd & \quad \yes \hfill \no \hfill \no \hfill \yes \mbox{\quad} &
  \quad \yes \hfill \no \hfill \yes \hfill \no \mbox{\quad} \\ \hline
\end{tabular}
\caption{Appearance of dangling bonds or Zeeman-field terms at the edges of the compass subsystems 1 (with odd $x$) and 2 (with even $x$), where \yes{} means that such terms appear and \no{} means that they do not. The boundary conditions are open in both directions.}
\label{tab.edgefields}
\end{table}

So far, we have not discussed the edges of the system \footnote{We do not discuss the cases of $L_x=2$ or $L_y=2$ here, in which the system consists only of edges. These cases are integrable \cite{chiuinteract} and thus nongeneric.}. As seen from Fig.\ \ref{fig.compass_map2}, there are always ``dangling bonds'' in the row $y=1$ as well as, for one of the compass subsystems, in the column $x=1$. Dangling bonds in the first row or column represent plaquette terms in this row or column, which are mapped onto only a single spin operator, as noted above. This corresponds to a static, uniform magnetic field applied at these edges.

\begin{table*}
\begin{tabular}{|c|ccc|ccc|} \hline
  & \multicolumn{3}{|c|}{subsystem 1} & \multicolumn{3}{|c|}{subsystem 2} \\
\raisebox{1.2ex}[1.2ex]{$L_x \times L_y$} & quantum & classical & static & quantum & classical & static \\ \hline\hline
\vstrut even $\times$ even & $\frac{L_x(L_y-2)}{4}$ & $\frac{L_x}{2}$ & $\frac{L_x}{2}$ &
$\frac{(L_x-2)(L_y-2)}{4}$ & $\frac{L_x-2}{2} + \frac{L_y-2}{2}$ & $\frac{L_x-2}{2} + \frac{L_y-2}{2}$ \\ \hline
\vstrut even $\times$ odd & $\frac{L_x(L_y-1)}{4}$ & $0$ & $\frac{L_x}{2}$ &
$\frac{(L_x-2)(L_y-1)}{4}$ & $\frac{L_y-1}{2}$ & $\frac{L_x-2}{2} + \frac{L_y-1}{2}$ \\ \hline
\vstrut odd $\times$ even & $\frac{(L_x-1)(L_y-2)}{4}$ & $\frac{L_x-1}{2}$ & $\frac{L_x-1}{2} + \frac{L_y-2}{2}$ &
$\frac{(L_x-1)(L_y-2)}{4}$ & $\frac{L_x-1}{2}$ & $\frac{L_x-1}{2} + \frac{L_y-2}{2}$ \\ \hline
\vstrut odd $\times$ odd & $\frac{(L_x-1)(L_y-1)}{4}$ & $0$ & $\frac{L_x-1}{2} + \frac{L_y-1}{2}$ &
$\frac{(L_x-1)(L_y-1)}{4}$ & $0$ & $\frac{L_x-1}{2} + \frac{L_y-1}{2}$ \\ \hline
\end{tabular} 
\caption{Enumeration of terms and degrees of freedom in the two compass subsystems 1 (with odd $x$) and 2 (with even $x$). The boundary conditions are open in both directions. For each subsystem, the number of quantum spins (for which both $\sigma^x$ and $\sigma^z$ appear in the Hamiltonian), of classical degrees of freedom at the edges (for which only one spin component appears), and of static-magnetic-field terms at the edges are given in consecutive columns.}
\label{tab.spincount1}
\end{table*}

\begin{table*}
\begin{tabular}{|c|cccc|c|} \hline
  & \multicolumn{4}{|c|}{number of degrees of freedom} & \\
\raisebox{1.2ex}[1.2ex]{$L_x \times L_y$} & Majorana & subsystem 1 & subsystem 2 & both & \raisebox{1.2ex}[1.2ex]{undercount} \\ \hline\hline
\vstrut even $\times$ even & $\frac{L_xL_y}{2}$ & $\frac{L_xL_y}{4}$ & $\frac{L_xL_y-4}{4}$ & $\frac{L_xL_y}{2}-1$ & $1$
  \\ \hline
\vstrut even $\times$ odd & $\frac{L_xL_y}{2}$ & $\frac{L_x(L_y-1)}{4}$ & $\frac{L_x(L_y-1)}{4}$ & $\frac{L_x(L_y-1)}{2}$ & $\frac{L_x}{2}$
  \\ \hline
\vstrut odd $\times$ even & $\frac{L_xL_y}{2}$ & $\frac{(L_x-1)L_y}{4}$ & $\frac{(L_x-1)L_y}{4}$ & $\frac{(L_x-1)L_y}{2}$ & $\frac{L_y}{2}$ \\ \hline
\vstrut odd $\times$ odd & $\frac{L_xL_y+1}{2}$ & $\frac{(L_x-1)(L_y-1)}{4}$ & $\frac{(L_x-1)(L_y-1)}{4}$ & $\frac{(L_x-1)(L_y-1)}{2}$ & $\frac{L_x+L_y}{2}$ \\ \hline
\end{tabular}
\caption{Numbers of degrees of freedom in the Majorana model (including a spectator mode for the odd times odd case) and the two compass models it is mapped onto, as well as the resulting total number for both subsystems. The latter is always smaller than the number of degrees of freedom of the Majorana model. The last column shows the difference.}
\label{tab.spincount2}
\end{table*}

Moreover, there can also be dangling bonds in the last row, $y=L_y-1$, or in the last column, $x=L_x-1$. Table \ref{tab.edgefields} summarizes at which edges dangling bonds appear. For dangling bonds in the last row or column, spin operators occur that represent the product of all Majorana operators in two adjacent columns or rows, namely~\footnote{The prefactors result from the combination of the factors in Eqs.\ (\ref{eq.cm.sigmax}) and (\ref{eq.cm.sigmaz}) and factors of $i$ included in the definitions of $R_y$ and $C_x$ for Majorana strings of certain lengths.}
\begin{align}
\sigma^x_{L_x,y} &= \left\{\begin{array}{cl}
    1 & \mbox{for $L_x$ even} \\
    i & \mbox{for $L_x$ odd}
  \end{array} \right\} \times R_y R_{y+1} ,
\label{eq.cm.sigmaxedge} \\
\sigma^z_{x,L_y} &= i^{L_y}\, C_x C_{x+1} .
\label{eq.cm.sigmazedge}
\end{align}
These products are compatible integrals of motion for any $L_x$ and $L_y$ \footnote{This statement seems to be contradicted by the case of $\sigma^x_{L_x,L_y}$ and $\sigma^z_{L_x,L_y}$ but these operators do not occur in the compass subsystems.}. Hence, the Hamiltonian can be block diagonalized with respect to all of these quantities, which can thus be interpreted as classical degrees of freedom. In each block, they appear as generally nonuniform magnetic fields acting on the last row or column.

It will prove useful to minimize the number of degrees of freedom. This is possible for subsystem 1 in the case of odd $L_x$, where subsystem 1 has dangling bonds in the last column ($x=L_x-1$) but not in the first. We can then redefine the operators $\sigma^x_{x,y}$ for subsystem 1 in terms of products starting from the right edge. The result is that now the right edge contains a static magnetic field and the classical degrees of freedom in Eq.\ (\ref{eq.cm.sigmaxedge}) do not appear. Moreover, the two subsystems are then equivalent and thus have the same spectrum. This trick is not useful for even $L_x$ since then subsystem 2 contains dangling bonds on both the left and right edges, whereas subsystem 1 does not contain dangling bonds at either edge. Table \ref{tab.spincount1} summarizes, for each of the two compass subsystems, the number of quantum spin-$1/2$ degrees of freedom (for which both the $x$ and the $z$ component appear), the number of classical degrees of freedom at the edges, and the number of constant magnetic-field terms acting at the edges.

It is instructive to compare the total number of degrees of freedom of the compass subsystems with the one of the original Majorana model. For this purpose, the static-magnetic-field terms do not count but the classical degrees of freedom do. Table \ref{tab.spincount2} lists the resulting numbers of degrees of freedom, as well as the difference between the Majorana and compass models. We see that the compass models always have a smaller number of degrees of freedom than the Majorana model. The difference must refer to degrees of freedom that do not appear in the Hamiltonian. The Hilbert space of the Majorana model is thus equal to the direct product of the Hilbert spaces of the two compass models times another Hilbert space of these extra degrees of freedom.

The dimension of the Hilbert space of the decoupled degrees of freedom is two to the power given in the last column of table \ref{tab.spincount2}. This implies that all eigenenergies have a degeneracy that is an integer multiple of this dimension. Intriguingly, this is exactly the ``topological'' degeneracy we have found based on the Clifford algebra, for all four cases. Numerically, we do not find any remaining degeneracy of the ground states of the two compass models (excited states may be degenerate due to broken reflection and rotation symmetries). In this sense, the mapping maximally simplifies the problem.

It should be pointed out that the absence of ground-state degeneracy of the compass subsystems is due to the edge terms that appear for all cases, see table \ref{tab.edgefields}. Without these, the compass models with open, periodic, or mixed boundary conditions show even degeneracy of all eigenstates and, in particular, twofold degeneracy of the ground state~\cite{Doucot2005, RevModPhys.87.1}.

The mapping to the two compass models is only possible for open boundary conditions in both directions. The reason is that the mapping is nonlocal, involving strings of Majorana operators reaching up to the plaquette in question, see Fig.\ \ref{fig.compass_map1}. These strings must start somewhere. One can of course connect opposite edges to obtain compass models with periodic boundary conditions in one or both directions. These models are not required to be equivalent to the Majorana models with the same boundary conditions. Indeed, exact diagonalization of Majorana and compass models with sizes up to $3\times 9$, $4\times 7$, and $5\times 6$ shows that the spectra coincide, including degeneracies, for open boundary conditions in both directions but not for any other case.

Kamiya \textit{et al.}\ \cite{PhysRevB.98.161409} also map the interacting Majorana model onto two decoupled quantum compass models. Since they are only interested in the thermodynamic limit they disregard any edge terms. It does not seem obvious to us that this is justified for a model with string invariants and we analyze the effect of edge terms below. In any case, we are also interested in finite systems and therefore must take the edges into account. This was also necessary to understand the global degeneracy of the spectrum in terms of decoupled degrees of freedom.

\begin{figure}[tb]
\includegraphics[scale=0.55]{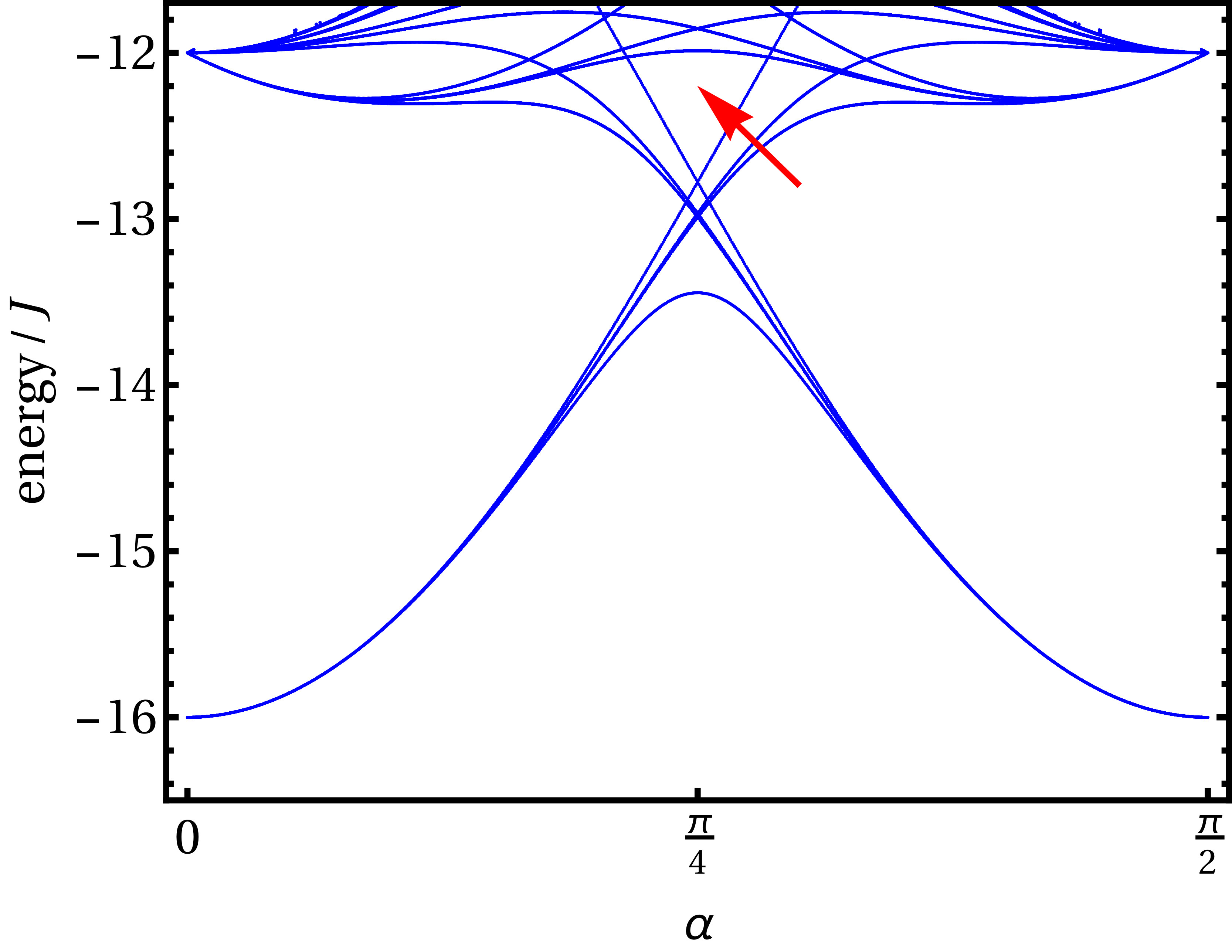}
\caption{Low-energy levels for the compass model with periodic boundary conditions and anisotropic couplings $J_x=J\cos\alpha$, $J_y=J\sin\alpha$ in Eq.\ (\ref{eq.Hcompass.ani}). The system size is $L=4$. The red arrow indicates the nascent gap above the lowest $2^{L+1}-2$ levels.}
\label{fig.compass_ani}
\end{figure}

The compass model on the square lattice without fields at the edges has been studied extensively \cite{RevModPhys.87.1}. The classical compass model with periodic boundary conditions exhibits a continuous ground-state degeneracy under uniform $\mathrm{SO}(2)$ rotations of the spins, besides additional invariances under discrete transformations \cite{RevModPhys.87.1}. For both the classical and the quantum compass model, the existence of one-dimensional gauge-like invariants, namely row and column operators, prevents spontaneous order of the spins \cite{PhysRevB.72.045137,RevModPhys.87.1}. In our notation, the row operators for the two subsystems $s=1,2$ are $R^1_y = \prod_{x\text{ odd}} \sigma^z_{x,y}$ and $R^2_y = \prod_{x\text{ even}} \sigma^z_{x,y}$, where $y$ is always even. The corresponding column invariants are $C^1_x = \prod_{y\text{ even}} \sigma^x_{x,y}$, where $x$ is odd, and $C^2_x = \prod_{y\text{ even}} \sigma^x_{x,y}$, where $x$ is even.

On the other hand, both the classical and the quantum compass models show a finite-temperature phase transition to a spin-nematic state \cite{PhysRevLett.93.207201, PhysRevB.72.024448, PhysRevB.78.064402, PhysRevB.87.214421, PhysRevB.98.161409, RevModPhys.87.1}. Its order parameter
\begin{equation}
\Delta = \big\langle \sigma^x_{x,y}\sigma^x_{x+2,y} - \sigma^z_{x,y}\sigma^z_{x,y+2} \big\rangle
\label{eq.Delta.nematic}
\end{equation}
is of Ising type. For the quantum model, the order is accompanied by an energy gap between a highly degenerate ground state and the excited states \cite{Doucot2005, PhysRevB.72.024448, PhysRevB.72.045137, PhysRevB.87.214421, RevModPhys.87.1}. For $L\times L$ spins and periodic boundary conditions, the ground-state degeneracy is exactly twofold but in the thermodynamic limit $2^{L+1}-2$ states approach the same ground-state energy.
These states develop out of the $2^L$ ground states of decoupled rows and the $2^L$ ground states of decoupled columns upon changing the row and column couplings in the Hamiltonian
\begin{equation}
H = \sum_{x,y} \left( J_x\, \sigma^x_{x,y} \sigma^x_{x+1,y} + J_y\, \sigma^z_{x,y} \sigma^z_{x,y+1} \right)
\label{eq.Hcompass.ani}
\end{equation}
independently. The doublet of uniform row or column states is common to both limits so that the total number is $2\times 2^L-2$. For illustration, the low-energy part of the spectrum for $L=4$ and periodic boundary conditions is shown as a function of the coupling anisotropy in Fig.\ \ref{fig.compass_ani}. The approach is thought to be exponential, i.e., the energy differences scale as $\mathcal{O}(e^{-L/L_0})$ \cite{PhysRevB.72.024448, PhysRevB.87.214421, RevModPhys.87.1}. There are conflicting statements as to whether another doublet also approaches the ground-state energy in the thermodynamic limit \cite{PhysRevB.72.024448, PhysRevB.87.214421}, which would lead to a degeneracy of $2^{L+1}$. In any case, the degeneracy is much larger than the twofold degeneracy expected for a broken Ising symmetry. This is linked to the existence of the gauge-like row and column invariants~\cite{PhysRevB.72.045137, RevModPhys.87.1}.

As we have shown, the mapping from the Majorana model only works for open boundary conditions and unavoidably introduces Zeeman-type terms at edges of the compass models. Both the boundary conditions as well as the Zeeman field act on a one-dimensional subset of sites and one would thus expect them to be irrelevant for the ordering in the thermodynamic limit. To check whether the sites in the bulk decouple from the edges in this limit, we consider the classical compass model, which allows us to study much larger systems. Such an approach has proved fruitful for the compass model with periodic boundary conditions \cite{RevModPhys.87.1}. The classical model is described by the Hamiltonian in Eq.\ (\ref{eq.H.compass.2}), which is now understood as a classical function of two-component unit vectors $(\sigma^x_{x,y},\sigma^z_{x,y})$.

\begin{figure}[thb]
\includegraphics[scale=0.69]{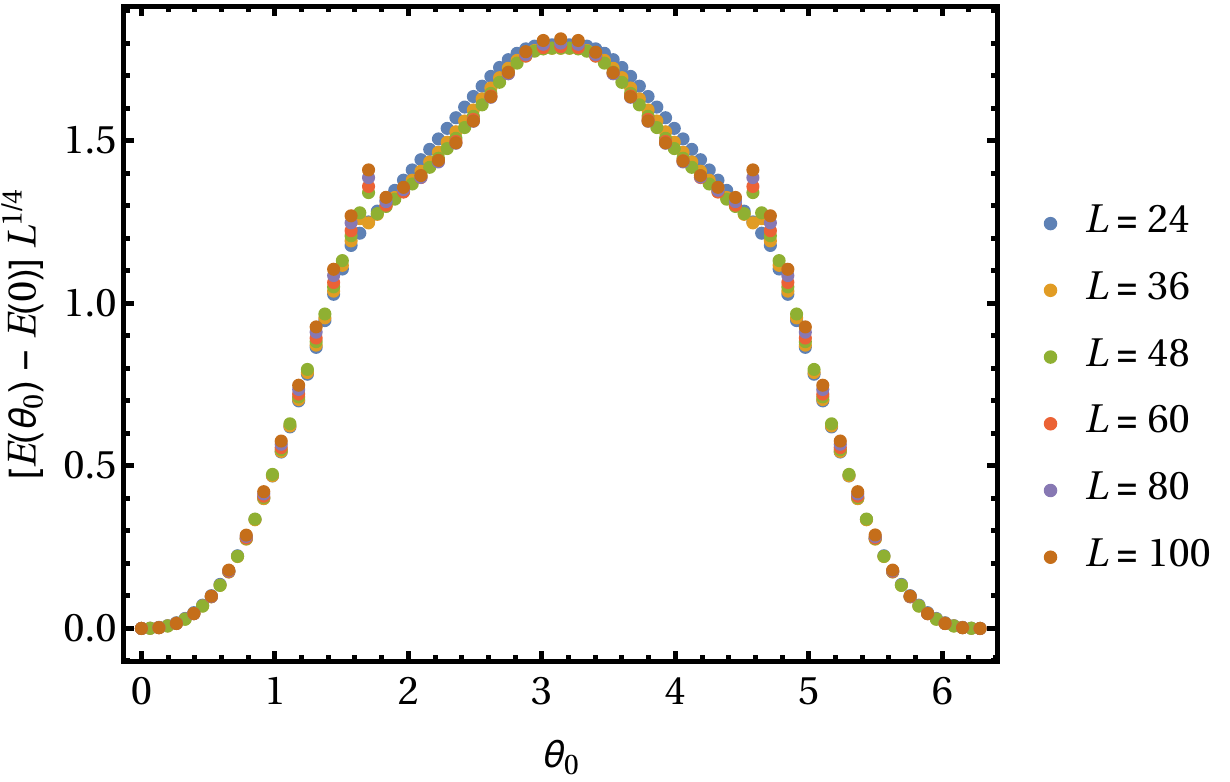}
\caption{Minimal energy $E(\theta_0)-E(0)$ for rotations of a spin at the center of the classical compass model on lattices of various sizes $L\times L$ with open boundary conditions and a magnetic field along the \textit{z}-direction applied at the bottom edge. $\theta_0=0$ corresponds to the center spin being parallel to the edge field. The energy is shown scaled with $L^{1/4}$.}
\label{fig.class_comp}
\end{figure}

As table \ref{tab.edgefields} shows, the simplest case is subsystem 1 for even times odd lattices, where there is a uniform magnetic field applied to the bottom row. We focus on this case in the following since more complicated edge terms should not affect our general conclusions. The coupling is assumed to be ferromagnetic, $g<0$. The antiferromagnetic model can me mapped onto the ferromagnetic one by rotating the spins on one checkerboard sublattice by $\pi$.
We find that the Zeeman field reduces the ground-state degeneracy to twofold. The two states show nonzero magnetization $\langle \sigma^z\rangle$ in the direction of the edge field in the thermodynamic limit. However, whether this magnetization survives at temperatures $T>0$ depends on the stiffness of magnetic excitations. We parametrize the classical spin vectors in terms of angles $\theta_{x,y}$ as
\begin{equation}
\sigma^x_{x,y} = \cos\theta_{x,y}, \quad \sigma^z_{x,y} = \sin\theta_{x,y} .
\end{equation}
Fixing one spin at the center by setting $\theta_{\lceil L/2\rceil,\lceil L/2\rceil} = \theta_0$ and adapting all other spins to minimize the energy, we obtain the energy cost $E(\theta_0)-E(\theta_0=0)$ of rotating the center spin, which is shown in Fig.\ \ref{fig.class_comp} for various system sizes. Evidently, $E(\theta_0)$ approaches zero for $L\to\infty$, meaning that rotations of the center spin become soft. Since the system is two dimensional we conclude that the magnetization $\langle \sigma^z\rangle$ vanishes in the thermodynamic limit.
The approach is very slow---the energy barrier for a rotation by $2\pi$ scales as $1/L^{1/4}$. This slow approach suggests that it is impossible to observe the decoupling of the bulk from the edge by exact diagonalization of the \emph{quantum} compass model with edge terms.

We now turn to the spin-nematic order. Since it is of Ising type it can occur at nonzero temperatures. The edge fields explicitly break the spin rotation symmetry and also lift the degeneracy between nematically ordered states with opposite order parameters $\Delta$. While, to our knowledge, the compass model with symmetry-breaking boundary terms has not been studied, we expect it to behave similarly to the Ising model with a symmetry-breaking boundary since the transitions of the unperturbed models belong to the same universality class.

It was shown that the partition function of the two-dimensional ferromagnetic Ising model with a magnetic field applied at one edge can, in the thermodynamic limit, be written as a sum of bulk and edge contributions \cite{PhysRev.162.436}. Hence, bulk and edge states decouple asymptotically. The magnetization as a function of the distance $y$ from the edge was studied in Refs.\ \cite{PhysRev.162.436, Abr80, AbH88}. If the spontaneous magnetization $m^*$ in the bulk is in the same direction as the magnetization induced by the edge field, the local magnetization $m(y)$ approaches $m^*$ exponentially fast as a function of $y$ for all temperatures below the Ising transition at $T_c$ \cite{AbH88}. On the other hand, if $m^*$ is in the opposite direction, $m(y)$ approaches $m^*$ exponentially fast only for $T<T_w<T_c$, where $T_w$ is the critical temperature of a \emph{wetting transition} \cite{Abr80,AbH88,ABH90,Mac96,VAM05}. At this transition, a domain wall separating regions with opposite sign of the magnetization decouples from the edge.


Our model is more complicated, however, since subsystem 2 always has boundary terms at two or more edges, see table \ref{tab.edgefields}. A systematic study of the possible wetting transition for the interacting Majorana model would be worthwhile. We conjecture that the Majorana model with open boundary conditions also shows a wetting transition at a temperature $T_w<T_c$ and now focus on the temperature range below $T_w$. Here, any effect of the edges decays exponentially into the bulk. In particular, in the thermodynamic limit, the bulk shows spontaneous symmetry breaking described by the nematic order parameter $\Delta$, accompanied by an energy gap. However, nothing precludes states localized at the edges to be present within this gap.

It is not obvious how many bulk states will end up below the gap and collapse to the ground state in the thermodynamic limit. In analogy to the corresponding asymptotic number $2^{L+1}$ for periodic boundary conditions, we can denote the asymptotic number of ground states by $2^{L_\text{eff}+1}$, where $L_\text{eff}$ is the effective linear dimension of the bulk region that decouples from the edges. The exponential decay of edge effects suggests that $L_\text{eff}/L$ approaches unity for $L\to\infty$.

From now on, we assume that the boundary conditions and the edge fields become irrelevant for the bulk of the Majorana system in the thermodynamic limit. We can then use compass models with periodic boundary conditions to infer results for the Majorana model. In this spirit, upon mapping back onto the Majorana model, the order in each of the two decoupled compass models corresponds to an antiferroic order of plaquette terms $\langle \zeta_{x,y} \zeta_{x+1,y} \zeta_{x+1,y+1} \zeta_{x,y+1} \rangle$ on each of the two checkerboard sublattices. This results in four distinct stripe orderings, as found in Ref.\ \cite{PhysRevB.98.161409}.
Interestingly, Eqs.\ (\ref{eq.cmmapx}) and (\ref{eq.cmmapz}) show that the nematic order parameter is local in both representations, although the mapping between Majorana and compass models is nonlocal. For the square lattice, the wavelength of the stripe order is fixed to twice the lattice constant, $\lambda = 2a = 2\, \sqrt{s_\text{uc}\, N/N_\text{flat}}$.

\subsection{Integrable ladder models}
\label{sub.ladder}

For open boundary conditions, the mapping to two decoupled compass models and additional spectator degrees of freedom reveals the integrability of a number of special cases. The two-leg Majorana ladder is trivially integrable since all plaquette terms commute~\cite{chiuinteract}. We note that two-leg and four-leg ladders with a bilinear term in the Hamiltonian and periodic boundary conditions have recently been studied by Rahmani \textit{et al.}~\cite{RPA19}.

The mapping shows that, in addition, the three-leg and four-leg ladders are integrable in the sense of Braak \cite{Bra11}. The three-leg ladder of even length $L_x$ maps onto two decoupled spin models with Hamiltonians
\begin{align}
H_1 &= g \sum_{x=1 \atop \text{odd}}^{L_x-3} \sigma^x_{x,2} \sigma^x_{x+2,2}
  + g \sum_{x=1 \atop \text{odd}}^{L_x-1} \sigma^z_{x,2} , \\
H_2 &= g \sum_{x=2 \atop \text{even}}^{L_x-2} \sigma^x_{x,2} \sigma^x_{x+2,2}
  + g\, \sigma^x_{2,2} + g \sum_{x=2 \atop \text{even}}^{L_x-2} \sigma^z_{x,2} ,
\end{align}
see also Fig.\ \ref{fig.compass_map2}. The first is a critical one-dimensional transverse-field Ising model, and the second in addition has a field in the longitudinal direction applied at one end and no field applied to the spin $\mbox{\boldmath$\sigma$}_{L_x,2}$ at the other end. The three-leg ladder of odd length $L_x$ maps onto two decoupled spin models described by Hamiltonians $H_s$, where
\begin{equation}
H_2 = g \sum_{x=2 \atop \text{even}}^{L_x-3} \sigma^x_{x,2} \sigma^x_{x+2,2}
  + g\, \sigma^x_{2,2} + g \sum_{x=2 \atop \text{even}}^{L_x-1} \sigma^z_{x,2}
\end{equation}
and $H_1$ is equivalent to $H_2$ upon redefining $\sigma^x_{x,y}$, see above. The two subsystems are critical one-dimensional transverse-field Ising models with an additional field in the longitudinal direction at one end. These spin models are integrable and can be solved by refermionization \cite{LSM61, PFEUTY197079, Campostrini_2015, Igloi_2008}. In this way, we have obtained the energy gap between the degenerate ground and first excited states:
\begin{equation}
E_1 - E_0 = 4g \times \left\{\begin{array}{ll}
    \displaystyle\sin\left( \frac{1}{2}\, \frac{\pi}{L_x+1} \right) & \mbox{for $L_x$ even} , \\[2.3ex]
    \displaystyle\sin\left( \frac{\pi}{L_x+1} \right) & \mbox{for $L_x$ odd} .
  \end{array}\right.
\label{eq.E1E0.3leg.2}
\end{equation}
Evidently, the gap scales as a power law of the length $L_x$ for large $L_x$, which can be attributed to the criticality of the transverse-field Ising models.

The four-leg ladder of even length $L_x$ maps onto spin models with Hamiltonians
\begin{align}
H_1 &= g \sum_{x=1 \atop \text{odd}}^{L_x-3} \sigma^x_{x,2} \sigma^x_{x+2,2}
  + g \sum_{x=1 \atop \text{odd}}^{L_x-1} \sigma^z_{x,2}\, ( 1 + \sigma^z_{x,4} ) , \\
H_2 &= g \sum_{x=2 \atop \text{even}}^{L_x-2} \sigma^x_{x,2} \sigma^x_{x+2,2}
  + g\, \sigma^x_{2,2}
  + g \sum_{x=2 \atop \text{even}}^{L_x-2} \sigma^z_{x,2}\, ( 1 + \sigma^z_{x,4} ) .
\end{align}
Here, $\mbox{\boldmath$\sigma$}_{x,4}$ are classical spins that commute with $H_s$. Thus $H_s$ can be block diagonalized with respect to these spins. Refermionization leads to $E_1 - E_0 = 2g\sqrt{5-4\cosh\nu}$ for even $L_x$, where $\nu$ is the solution of the equation~\cite{LSM61}
\begin{equation}
\frac{\sinh\left( \left(\frac{L_x}{2}+1\right) \nu \right)}{\sinh\left( \frac{L_x}{2}\, \nu \right)} = 2 ,
\end{equation}
which can be solved numerically. For large $L_x$, the energy gap can be expanded as~\cite{Pfe79}
\begin{equation}
\frac{E_1 - E_0}{g} \cong \frac{1}{1 - \frac{1}{2^{L_x-2}}}\, \frac{3}{2^{L_x/2}} .
\end{equation}
Unlike for the three-leg ladder, the gap closes exponentially for increasing length $L_x$.

The four-leg ladder with odd $L_x$ maps onto two spin models with
\begin{equation}
H_2 = g \sum_{x=2 \atop \text{even}}^{L_x-3} \sigma^x_{x,2} \sigma^x_{x+2,2}
  + g\, \sigma^x_{2,2}
  + g \sum_{x=2 \atop \text{even}}^{L_x-1} \sigma^z_{x,2}\, ( 1 + \sigma^z_{x,4} )
\end{equation}
and $H_1$ equivalent to $H_2$ upon redefining $\sigma^x_{x,y}$. Refermionization shows that the smallest excitation energy remains finite for $L_x\to\infty$. The gap is determined by the $(L_x+1)\times(L_x+1)$ matrices~\cite{ITK97}
\begin{equation}
T(s) = \left(\begin{array}{cccccccccc}
0 & 2 & & & & & & & & \\
2 & 0 & 1 & & & & & & & \\
 & 1 & 0 & 2 & & & & & & \\
 & & 2 & 0 & & & & & & \\
 & & & & \ddots & & & & & \\
 & & & & & 0 & 1 & & & \\
 & & & & & 1 & 0 & s & & \\
 & & & & & & s & 0 & 1 & \\
 & & & & & & & 1 & 0 & 0 \\
 & & & & & & & & 0 & 0
\end{array}\right) .
\end{equation}
The energy gap is
\begin{equation}
E_1 - E_0 = g \! \sum_{n=1}^{(L_x+1)/2} [\lambda(2)_n - \lambda(0)_n] ,
\end{equation}
where the $\lambda(s)_n$ are the positive eigenvalues of the matrix $T(s)$ \cite{ITK97}. Finding these eigenvalues still requires \emph{exponentially} smaller resources than diagonalizing the Hamiltonians $H_s$. For $L_x\to\infty$, the gap approaches
\begin{equation}
\frac{E_1 - E_0}{g} \cong \frac{6}{\pi}\, E\bigg(\frac{\pi}{2} \bigg| \frac{8}{9}\bigg) - 1 ,
\end{equation}
where
\begin{equation}
E(\phi|m) = \int_0^\phi d\theta\, \sqrt{1 - m \sin^2\theta}
\end{equation}
is the incomplete elliptic integral of the second kind. For the four-leg ladder with odd $L_x$, the gap corresponds to flipping a classical spin at the end of the ladder, unlike for all other cases, where the lowest-energy excitation is a fermionic one of the refermionized model.

Although the results for $E_1-E_0$ for even or odd length are very different, the results for the full spectra are in fact quite similar. For even length, we have found low-lying excited states that for increasing length approach the ground state exponentially fast. For odd length, the corresponding states are always part of the ground-state manifold and are thus not reflected in $E_1-E_0$. For both even and odd lengths, there is a gap of order $g$ above the low-energy sector.

\subsection{Spectrum and energy gap}
\label{sub.spectrum}

The mapping can also be exploited for numerical exact diagonalization of the Majorana model. The dimension of the Hilbert space of the larger compass subsystem is $2^{\lfloor L_x/2 \rfloor \lfloor L_y/2 \rfloor}$ in all cases. Numerical efficiency is further increased by block diagonalizing the compass models in the presence of classical degrees of freedom.

\begin{figure}[thb]
\raisebox{2ex}[2ex]{(a)\hspace{-1em}}\includegraphics[scale=0.67]{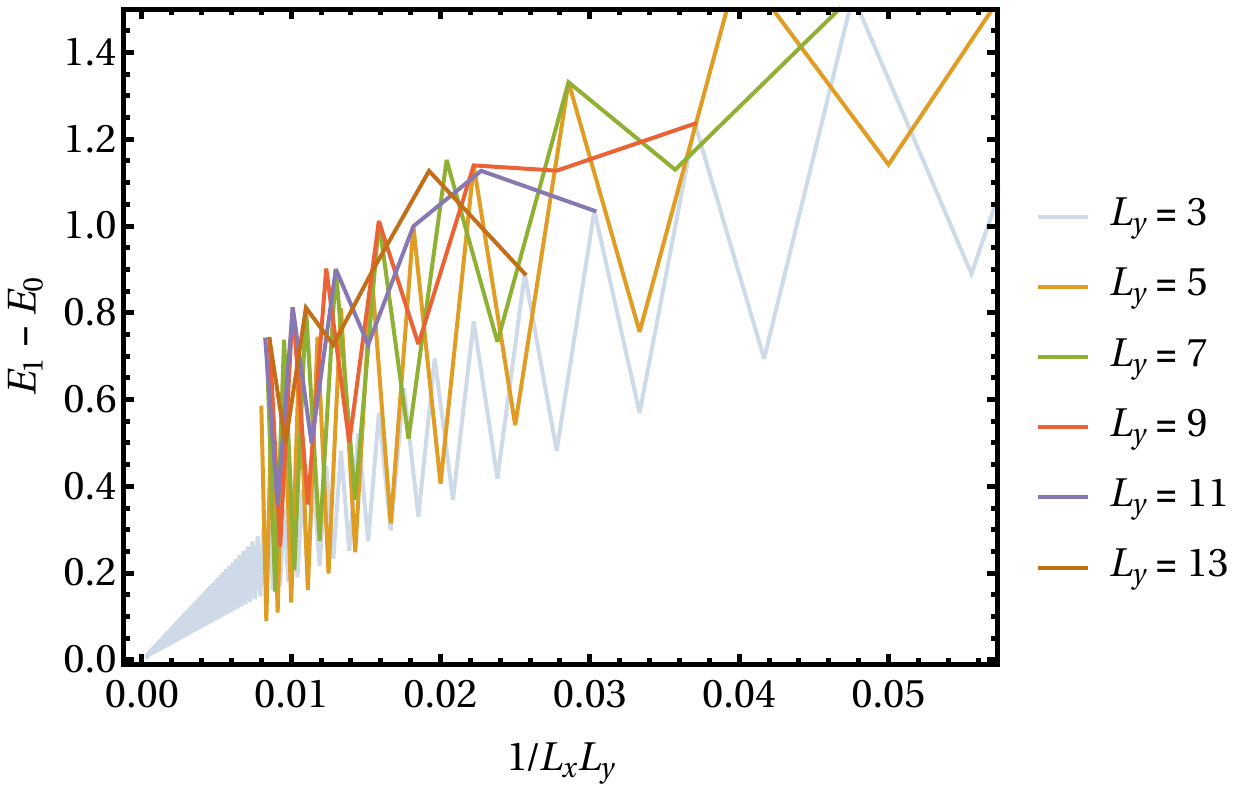}\\[2ex]%
\raisebox{2ex}[2ex]{(b)\hspace{-1em}}\includegraphics[scale=0.67]{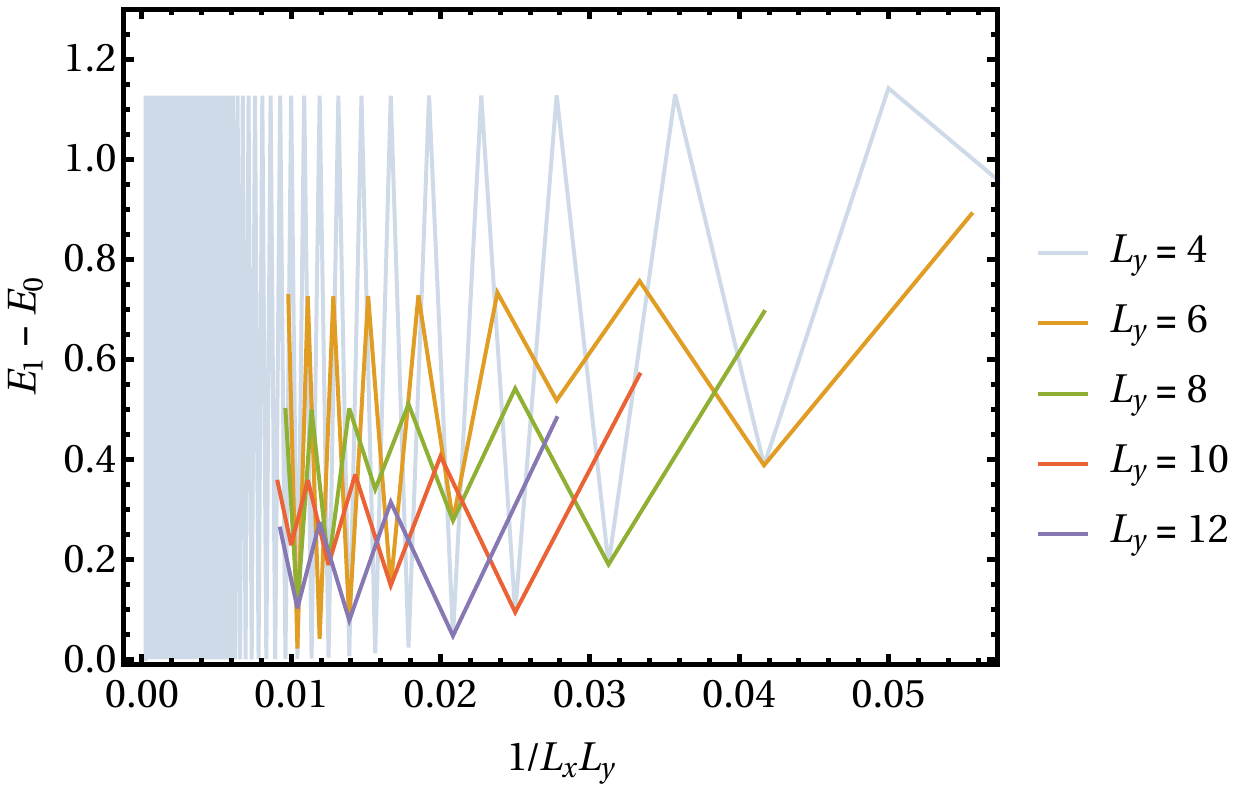}
\caption{Lowest excitation energy $E_1-E_0$ of interacting Majorana models as a function of the inverse area, $1/L_xL_y$, for $L_y$ (a) odd and (b) even. Data points for equal $L_y$ are connected. The data were obtained by exact diagonalization of compass Hamiltonians, except for the integrable cases $L_x=3,4$, where refermionization was employed.}
\label{fig.deltaEopen}
\end{figure}

As an application, we study the energy gap $E_1-E_0$ between the degenerate ground and first excited states, plotted in Fig.\ \ref{fig.deltaEopen} for system sizes up to $25\times 5$, $17\times 7$, $13\times 9$, and $11 \times 11$. For $L_y=3,4$, Fig.\ \ref{fig.deltaEopen} shows results obtained using refermionization, as discussed in Sec.\ \ref{sub.ladder}; the results agree with exact diagonalization up to sizes of $49\times 3$ and $25\times 4$.

For odd widths $L_y$, the gap closes for increasing length $L_x$. For $L_y=3$, we have seen in Sec.\ \ref{sub.ladder} that asymptotically $E_1-E_0$ is a power law of $1/L_x$ with exponent $1$. For $L_y>3$, the numerical results are still consistent with power laws but the exponent is clearly larger (smaller) than unity for even (odd) lengths.

For even widths $L_y$ and odd length $L_x$, the gap remains open in the limit $L_x \to \infty$, as seen above for $L_y=4$. The results for even length are unexpected, though. While for widths of $L_y=6,8$ the gap might still close, it actually increases as a function of even $L_x$ for $L_y=10,12$.
In any case, the asymptotic gap approaches zero when we first take $L_x\to \infty$ and then $L_y\to \infty$ even, hence the gap vanishes in the two-dimensional thermodynamic limit. This is consistent with the collapse of exponentially many energy levels onto the ground state expected for the compass models.

For periodic and mixed boundary conditions, the mapping onto compass models is not possible. To calculate the spectrum one can form complex fermions out of pairs of Majorana modes, leading to a matrix representation of the Hamiltonian of dimension $2^{N_{\text{flat}}/2}$. Chiu \textit{et al}.\ \cite{chiuinteract} have obtained low-lying eigenenergies for large systems with $L_y=4$ held fixed. They find that the excitation energy approaches zero for $L_x \to \infty$, regardless of whether $L_x$ is even or odd, unlike for open boundary conditions. Our numerical results for periodic boundary conditions (not shown) agree with this result.

\subsection{Consequences for topological order}

We now return to the discussion of topological properties of the Majorana model, first focusing on even times even lattices. The row operators $R_y$ and column operators $C_x$ can be understood as \emph{string operators} in the sense of Refs.\ \cite{PhysRevB.67.245316, wenquantumorder, booklinkedcluster, Hamma, PhysRevB.72.045137}. The related toric-code model exhibits closed-string condensation \cite{wenquantumorder, booklinkedcluster, Hamma}: String operators with closed strings commute with the Hamiltonian so that the ground states can be chosen to be eigenstates of these string operators. On the other hand, open strings do not commute with the Hamiltonian. This is characteristic of topological order.

While for the toric code all closed strings commute with the Hamiltonian, in our model only straight string operators, i.e., $R_y$ and $C_x$, and their products do so. Hence, the ground states (and all eigenstates) of our model contain only strings that wrap around the system but do not contain any local strings, unlike for the toric code. The eigenvalues of the string operators can thus be interpreted as $\mathbb{Z}_2$ fluxes through the toroidal system for periodic boundary conditions, see Fig.~\ref{fig.torusflux}.

The algebraic properties of the string operators in our model are also distinct from the toric code: The operators $R_y$, $C_x$, and $\Gamma_{xy}=iC_xR_y$ for arbitrary but fixed $x$, $y$ satisfy the algebra of the Pauli matrices, i.e., describe a pseudospin $1/2$. The horizontal and vertical fluxes in Fig.\ \ref{fig.torusflux} are thus incompatible observables. Moreover, the spectrum consists of pseudospin doublets and this twofold degeneracy is robust to randomness of the plaquette couplings. Since all $R_y$ commute among themselves (as do the $C_x$) the ground-state subspace is spanned by two eigenstates of all $R_y$. These two eigenstates are macroscopically distinct in that all $R_y$ eigenvalues are reversed between them. This is seen as follows: Take $|\psi\rangle$ to be one of the ground states, with $R_y |\psi\rangle = r_y |\psi\rangle$ for all $y$ and $r_y=\pm 1$. Then $C_x|\psi\rangle$ satisfies $R_y C_x |\psi\rangle = - C_x R_y |\psi\rangle = - r_y C_x |\psi\rangle$. Since $C_x$ commutes with $H$ and $C_x|\psi\rangle$ has opposite $R_y$ eigenvalues compared to $|\psi\rangle$, $C_x|\psi\rangle$ must be the other member of the ground-state doublet.

The preceding argument works for any $C_x$, hence two operators $C_x$, $C_{x'}$ for arbitrarily distant columns perform the same mapping of one of the ground states onto the other. This implies that the ground states are long-range entangled. Of course, rows and columns can be interchanged in the preceding arguments.

To summarize, the interacting Majorana model with even times even dimensions has two macroscopically distinct, long-range entangled ground states that differ in $\mathbb{Z}_2$ fluxes through the system when put onto a torus. The twofold degeneracy is robust against random perturbations of the plaquette couplings. We conclude that the ground states show topological order of essentially the same type as the toric-code model. However, in the case of the toric code, the topological properties are robust against \emph{any} weak perturbation since they are protected by an energy gap \cite{PhysRevB.40.7133, KITAEVquantumcomputing}. This gap is due to the integrability of the toric-code model and is absent for our model, as shown in Sec.~\ref{sub.spectrum}.

On the other hand, at least for open boundary conditions, our model maps onto two compass models that show spontaneous symmetry breaking and a gap for bulk excitations, see Sec.\ \ref{sub.mapping}. 
The topological order could still be robust if only states with the same $\mathbb{Z}_2$ fluxes were present below the Ising gap. However, this is not the case, as we show in the following.

We assume, like in Sec.\ \ref{sub.mapping}, that at sufficiently low temperatures the effects of the edges are localized so that the bulk of the system can be analyzed without regarding the edges.
The model with even dimensions $L_x=L_y$ maps onto two compass models and a single decoupled mode. Each compass subsystem has linear dimension $L=L_x/2$ and on the order of $2^{L_x/2+1}$ states below the Ising gap. $2^{L_x/2}$ of these states transform into eigenstates for decoupled rows and $2^{L_x/2}$ transform into eigenstates of decoupled columns as the row and column couplings in the compass model are varied, see Eq.\ (\ref{eq.Hcompass.ani}). The eigenstates for decoupled rows are of course also eigenstates of all row operators $R^s_y$. The row operators remain invariants when the vertical coupling $J_y$ is switched on and thus the $2^{L_x/2}$ states can be chosen to be eigenstates of $R^s_y$ for all $J_y$. The corresponding eigenvalues are continuous functions of the couplings and, since they are integers, have to be constant. Now the limit of decoupled rows is trivially integrable and the eigenenergy of a single row does not change if the eigenvalue of the corresponding row operator $R^s_y$ is inverted. Consequently, all $2^L = 2^{L_x/2}$ combinations of eigenvalues occur in the ground-state sector for decoupled rows and, by continuity, also for isotropic coupling. Analogous arguments can be made involving the limit of decoupled columns. Hence, the ground-state sector of the Majorana model contains on the order of $2\times 2\times 2^{L_x/2+1} = 2^{L_x/2+3}$ states with any possible combination of eigenvalues of row or column invariants, i.e., of $\mathbb{Z}_2$ fluxes.

For odd $L_x=L_y$, an analogous argument also leads to the conclusion that states with any possible combination of row or column invariants occur in the ground-state sector. We conclude that for any $L_x=L_y$, and likely for all $L_x$, $L_y$, the Ising gap does not protect the $\mathbb{Z}_2$ fluxes against weak perturbations.

\section{Summary and conclusions}
\label{sec.summary}


The effects of interactions on the Majorana zero modes forming flat bands of surface states of topological NCSs have been analyzed. We have constructed a model for these modes by neglecting other low-energy excitations, e.g., in the bulk, and truncating the Hamiltonian after the leading interaction term. The Hamiltonian is then purely quartic since the kinetic energy is zero. The vanishing of the bilinear term is topologically protected by the winding numbers of line nodes in the bulk.

It is now crucial to realize that since Majorana modes exist in a subset of nonzero measure of the two-di\-men\-sio\-nal surface Brillouin zone, we can construct Majorana wave packets localized in real space that are nevertheless eigenstates of the Hamiltonian. This is quite remarkable as these wave packets do not disperse in the absence of perturbations. On the other hand, one could use external perturbations to manipulate them. NCS flat bands realize a new platform for interacting Majorana modes in two dimensions that does not require fine tuning of the chemical potential, unlike schemes involving Majorana modes bound to vortices \cite{Volovik1999, PhysRevLett.100.096407, chiuinteract, PhysRevB.98.161409, ARP17}.

Notably, any choice of real-space positions is possible as long as they have the correct density. For any choice, an orthonormal set of Majorana operators in real space needs to be constructed. We have done this using L\"owdin orthonormalization \cite{Loewdin}, which optimizes the localization of the orthonomalized wave packets.


Based on this groundwork, we have formulated a minimal model with plaquette interaction on a square lattice. It differs from the toric-code model \cite{PhysRevB.40.7133, KITAEVquantumcomputing} in that the plaquette terms do not all commute but anticommute if they share a corner, which makes our model nonintegrable.
In fact, the model has a full set of compatible integrals of motion but nevertheless is not integrable since these invariants are not independent.


The minimal model with any type of boundary conditions has a large number of string-like integrals of motion. Maximal anticommuting sets of these invariants form Clifford algebras, which imply degeneracies of all states that strongly depend on whether the linear dimensions $L_x$ and $L_y$ are even or odd. These degeneracies can be understood as topological since they persist for random plaquette couplings.
It is an interesting question for the future whether this even-odd dichotomy has observable consequences in the thermodynamic limit.


Furthermore, we have constructed a direct mapping of the interacting Majorana model onto two decoupled compass models and decoupled degrees of freedom, which is more transparent than the three-step mapping in Ref.\ \cite{PhysRevB.98.161409}. This mapping not only reduces the effective size by roughly one half but also divides out exactly the number of decoupled degrees of freedom that corresponds to the topological degeneracy. It thus maximally simplifies the problem of exact diagonalization. The mapping only works for open boundary conditions in both directions. The compass models obtained by the mapping necessarily contain edge terms, which strongly affect diagonalization results for feasible system sizes.


As two examples that profit from the mapping, we have shown the integrability of Majorana ladders with three and four legs and have studied the energy gap above the ground state for finite systems of sizes up to $11\times 11$ with open boundary conditions by exact diagonalization. If one dimension, say $L_y$, is held fixed at an odd value while the other, $L_x$, is send to infinity, this gap closes. On the other hand, if the fixed dimension $L_y$ is even, the gap remains open for odd $L_x\to \infty$, while the asymptotic behavior for even $L_x\to \infty$ depends on the width $L_y$. It closes exponentially for $L_y=4$. The results for $L_y=3$ and $4$ can be understood rigorously based on the integrability of these ladder models.


The type of boundary conditions should become irrelevant in the thermodynamic limit, at least at sufficiently low temperatures. The compass models may show a wetting transition but at low temperatures the perturbation by the boundaries should decay exponentially into the bulk. This is supported by calculations for the classical version of the model up to large system sizes, which show that the spins in the bulk decouple from the edges in the thermodynamic limit. Under this condition, the interacting Majorana model inherits the finite-temperature conventional (i.e., not topological) long-range order from the compass model \cite{PhysRevLett.93.207201, PhysRevB.72.024448, PhysRevB.78.064402, PhysRevB.87.214421, PhysRevB.98.161409, RevModPhys.87.1}. The Majorana model then shows a stripe-like modulation of the average $\langle \zeta_{x,y} \zeta_{x+1,y} \zeta_{x+1,y+1} \zeta_{x,y+1} \rangle$ with a wavelength of twice the lattice constant~\cite{PhysRevB.98.161409}.


These conclusions apply to Majorana modes on a square lattice with only plaquette interaction. If we take seriously that the real-space model is to represent the interacting zero-energy Majorana modes in $\mathcal{F}_\text{flat}$, we have to recall that we were free to choose the real-space lattice of localized Majorana modes. The transformation to a lattice generates also longer-range couplings $g_{ijkl}$ and these couplings are functions of the parameters of the chosen lattice. The symmetry breaking by the nematic order will happen in such a way that the free energy is minimized. This will fix the optimum lattice parameters together with the nematic order parameter if there is one. A prediction of the equilibrium state thus requires the calculation of the full coupling tensor $g_{ijkl}$ and subsequently of the free energy as functions of the lattice parameters, which is beyond the scope of this work.


The interacting Majorana model with even times even dimensions has two macroscopically distinct, long-range entangled ground states that differ in $\mathbb{Z}_2$ fluxes through the system when put onto a torus. The ground states show string condensation similar to the toric-code model. The degeneracy and the string condensation are robust against any random perturbations of the plaquette couplings. The ground states hence show topological order similar to the toric-code model. However, while the spectrum of bulk states develops a gap above the degenerate ground state in the thermodynamic limit due to the Ising-type conventional order, states with different values of the string invariants end up in the ground-state sector. Hence, the Ising gap does not protect the string invariants against weak perturbations and the topological order is not robust.
The Majorana model thus represents an interesting case of a nonintegrable system that is gapped and possesses fragile topological invariants. In view of the restriction to a short-range interaction in our model, it will be important to ascertain whether these are generic features of flat surface bands of noncentrosymmetric superconductors.

\acknowledgements

The authors thank P.\,M.\,R. Brydon, C.-K. Chiu, and S. Rex for useful discussions. Financial support by the Deut\-sche For\-schungs\-ge\-mein\-schaft through the Research Training Group GRK 1621, the Cluster of Excellence on Complexity and Topology in Quantum Matter ct.qmat (EXC 2147), and the Collaborative Research Center SFB 1143, project A04, is gratefully acknowledged.

\bibliography{Ruckert}

\end{document}